\documentclass{aastex7}
\graphicspath{{./}{images/}}

\begin{document}

\title{The Daniel K. Inouye Solar Telescope captures ambient shock waves ignited by magnetic vortices}

\author{Catherine E. Fischer}

\affiliation{European Space Agency (ESA), European Space Astronomy Center (ESAC),
Camino Bajo del Castillo s/n, 28692 Villanueva de la Ca{\~{n}}ada, Madrid, Spain
}
\affiliation{National Solar Observatory, Boulder, 3665 Discovery Drive, 80303 Boulder, USA}
\email[show]{}  

\author{Friedrich W{\"{o}}ger}
\affiliation{National Solar Observatory, Boulder, 3665 Discovery Drive, 80303 Boulder, USA}
\email{}  

\author{Gangadharan Vigeesh}
\affiliation{Kiepenheuer-Institute for Solar Physics (KIS), Georges-K{\"{o}}hler-Allee 401a, 79110 Freiburg, Germany}
\email{} 

\author{Peter H. Keys}
\affiliation{Astrophysics Research Center, School of Mathematics and Physics, Queen’s University Belfast, Belfast, BT7 1NN, UK}
\email{}

\author{Oskar Steiner}
\affiliation{Kiepenheuer-Institute for Solar Physics (KIS), Georges-K{\"{o}}hler-Allee 401a, 79110 Freiburg, Germany}
\affiliation{Istituto ricerche solari Aldo e Cele Dacc{\`{o}} (IRSOL), Faculty of Informatics, Universit{\`{a}} della Svizzera Italiana (USI), Via Patocchi 57, 6605 Locarno, Switzerland}
\email{}

\author{Jos\'e Roberto Canivete Cuissa}
\affiliation{Istituto ricerche solari Aldo e Cele Dacc{\`{o}} (IRSOL), Faculty of Informatics, Universit{\`{a}} della Svizzera Italiana (USI), Via Patocchi 57, 6605 Locarno, Switzerland}
\affiliation{Center for Theoretical Astrophysics and Cosmology, Institute for Computational Science (ICS), University of Zurich; Winterthurerstrasse 190, 8057 Z{\"{u}}rich, Switzerland.}
\email{}

\author{Shahin Jafarzadeh}
\affiliation{Astrophysics Research Center, School of Mathematics and Physics, Queen’s University Belfast, Belfast, BT7 1NN, UK}
\affiliation{Rosseland Centre for Solar Physics, University of Oslo, P.O. Box 1029, Blindern, 0315 Oslo, Norway}
\email{} 

\author{Robertus Erd{\'{e}}lyi}
\affiliation{Solar Physics \& Space Plasma Research Center (SP2RC), School of Mathematical and Physical Sciences, University of Sheffield, Hounsfield Road, Sheffield S3 7RH, UK}
\affiliation{Department of Astronomy, E{\"{o}}tv{\"{o}}s Lor{\'{a}}nd University, P{\'{a}}zm{\'{a}}ny P{\'{e}}ter s{\'{e}}t{\'{a}}ny 1/A, H-1112 Budapest, Hungary}
\affiliation{Hungarian Solar Physics Foundation, Pet{\H{o}}fit{\'{e}}r 3, H-5700 Gyula, Hungary}
\email{}

\author{David B. Jess}
\affiliation{Astrophysics Research Center, School of Mathematics and Physics, Queen’s University Belfast, Belfast, BT7 1NN, UK}
\affiliation{Department of Physics and Astronomy, California State University Northridge, Northridge, CA 91330, USA}
\email{}

\author{Mihalis Mathioudakis}
\affiliation{Astrophysics Research Center, School of Mathematics and Physics, Queen’s University Belfast, Belfast, BT7 1NN, UK}
\email{}

\author{Alexandra Tritschler}
\affiliation{National Solar Observatory, Boulder, 3665 Discovery Drive, 80303 Boulder, USA}
\email{}

\author{Damian J. Christian}
\affiliation{Department of Physics and Astronomy, California State University Northridge, Northridge, CA 91330, USA}
\email{}

\author{Andre Fehlmann}
\affiliation{National Solar Observatory, Boulder, 3665 Discovery Drive, 80303 Boulder, USA}
\affiliation{National Solar Observatory, 22 Ohi’a Ku Street, Makawao, HI 96768, USA}
\email{}

\author{Sarah Jaeggli}
\affiliation{National Solar Observatory, Boulder, 3665 Discovery Drive, 80303 Boulder, USA}
\affiliation{National Solar Observatory, 22 Ohi’a Ku Street, Makawao, HI 96768, USA}
\email{}

\author{Thomas Rimmele}
\affiliation{National Solar Observatory, Boulder, 3665 Discovery Drive, 80303 Boulder, USA}
\email{} 


\begin{abstract}

One of the first high-cadence chromospheric image sequences recorded with the recently commissioned Daniel K. Inouye Solar Telescope (DKIST) has revealed propagating arc-shaped bright fronts of a few hundred kilometers width. These bright fronts maintain their arc shape over up to one minute, travel with apparent horizontal speeds of $\sim$ $8$\,km s$^{-1}$, and originate from chromospheric bright grains seen at the center of the arc. We identify photospheric G-band bright points underneath the chromospheric bright grains, which are typically the signature of small-scale magnetic flux concentrations. The G-band bright points are found to interact with co-located vortices prior to the appearance of the chromospheric bright fronts. Images synthesized from three-dimensional magnetohydrodynamic simulations show a striking similarity with the discovered phenomenon and reveal that the arc-shaped bright fronts arise from essentially acoustic, weak shock fronts that propagate in the ambient medium, seemingly triggered by the vortex dynamics of the underlying magnetic flux concentration. In this work, we provide first hints of a hitherto unknown magnetohydrodynamic process by which vortex-driven torsional Alfvén pulses couple to predominantly acoustic waves that steepen into shocks and dissipate their energies. The detected number density and frequency of these events are at least a factor 10 too small to be a major source of solar chromospheric heating but simulations show a seven times higher number density. We therefore expect to discover many more events with tailored narrow band observations. 

\end{abstract}




\section{Introduction} \label{sec:intro}

Past studies of the Sun, currently the only highly resolvable star, have taught us that the discovery and understanding of the physical mechanisms at work in the solar atmosphere require observations of ever higher spatial and temporal resolution. 
By and by, such high-resolution observations revealed that the solar surface is abundantly occupied by small-scale magnetic flux concentrations at any given time (\citet{1973SoPh...33..281D,1996ApJ...463..365B,2004A&A...428..613B,2008ApJ...688.1374T, 2019MNRAS.488L..53K}; see also \citet{2009SSRv..144..275D} or \citet{2019LRSP...16....1B} for reviews). These magnetic flux concentrations reach superequipartition field strengths of $\approx 0.15$~T, presumably as a consequence of the convective collapse instability \citep{1979SoPh...61..363S,2008ApJ...677L.145N,2009A&A...504..583F}. Furthermore, high temporal resolution of image series of the solar atmosphere as seen, for example, in the visible continuum or the G-band, show a very dynamic solar atmosphere, changing on timescales of minutes and seconds \citep{2026ApJ...999..201K}. Magnetic field tracers, such as G-band bright points (BPs), are seen to be shuffled about by convective plasma motions, leading to a range of transient events. These (BPs) are a reliable indicator of small-scale magnetic flux concentrations \citep{2001ApJ...553..449B}, appearing bright due to the interplay between opacity effects, the magnetic field, and the surrounding dynamic plasma \citep{1976SoPh...50..269S,1996ApJ...463..797T,2001ASPC..236..445R,2001A&A...372L..13S}.  
More recently, a variety of small-scale processes involving such magnetic elements have been observed, such as (\textit{i}) dynamic fibrils \citep{2006ApJ...647L..73H} as slow-mode magnetoacoustic shocks propagating in magnetic conduits, (\textit{ii}) magnetic portals through which low-frequency ($<5$\,mHz) magnetoacoustic waves can propagate into the solar chromosphere \citep{2006ApJ...648L.151J}, (\textit{iii}) vortices in the chromosphere induced and driven by photospheric magnetic elements \citep{2012Natur.486..505W}, (\textit{iv}) torsional Alfv\'en pulses running along magnetic flux concentrations from the photosphere to the chromosphere \citep{2019NatCo..10.3504L}, or (\textit{v}) other oscillatory phenomena and jets such as those found by \citet{2017A&A...602L..12F} and \citet{2024A&A...691A..37D}, all linking the solar atmosphere across different physical regimes.

Studies of this connectivity between the solar atmospheric layers are highly sought after as they offer insight to the enigmatic heating problem\,---\,the fact that the chromosphere and corona display temperatures far exceeding the temperatures in the underlying photosphere.
Multi-dimensional numerical simulations have long predicted that photospheric drivers can excite magnetoacoustic waves and magnetic twist \citep{1998ApJ...495..468S,2011ApJ...727...17F, 2012ApJ...751L..21K, 2012A&A...541A..68M,2012ApJ...755...18V,2019ApJ...883..179K}. 
The waves and twist produced by the near-surface convection can propagate upward in the solar atmosphere and subsequently deposit their energies, possibly by developing into shocks or by other dissipation mechanisms \citep{1991A&A...241..625U,1999ApJ...519..899H,2007Sci...318.1574D,2007Sci...318.1572E,2009Sci...323.1582J,2012Natur.486..505W,2023LRSP...20....1J}. In fact, acoustic shocks are observed as periodic brightenings in the \ion{Ca}{2}\,H and K lines of the quiet Sun \citep{1997ApJ...481..500C} and are thought to be the dominant contributor in the heating of the lower chromosphere \citep{2019ARA&A..57..189C,  
2026A&A...705A..86N}. However, \citet{2005Natur.435..919F} came to the conclusion that high-frequency acoustic waves are not sufficient to heat the solar chromosphere. They inferred this from their analysis of the intensity fluctuation in the ultraviolet range of the solar spectrum as observed from space. On the other hand, the ubiquitous magnetic twist and associated torsional Alfv\'en waves are hard to dissipate although numerical simulations reveal waves near the boundaries of chromospheric magnetic vortices, which are likely damped by viscous heating \citep{2025A&A...700A.276C, 2026A&A...705A..42C}.

The National Science Foundation's Daniel K. Inouye Solar Telescope \citep[DKIST;][]{2020SoPh..295..172R}, currently the largest solar telescope in the world, has been designed for exactly this purpose: to study the Sun at unprecedented spatial resolution. DKIST is equipped with advanced adaptive optics technology and a diverse set of instruments for multiwavelength capabilities and it has now delivered the first science data sets during the currently ongoing Operations Commissioning Phase. In this work, we report on one of DKIST’s first discoveries, namely the presence of small-scale chromospheric wave fronts over the entire field-of-view (FoV), propagating for several hundred kilometers. They are seen to emanate from bright chromospheric grains that coincide with photospheric bright points, both tracers of the omnipresent small-scale magnetic flux concentrations \citep{2013A&A...549A.116J}.

With the aid of self-consistent, realistic numerical simulations, which show events very similar to the observed ones, we obtain valuable context for studying the underlying physical properties of these wave fronts, which turn out to be predominantly acoustic, weak shock fronts, triggered by torsional Alfvén pulses. We provide estimates on the occurrence rate and the potential of solar atmospheric heating by such events and point to the broader significance of this novel process.

The outline of the paper is as follows: in Section~\ref{sec:data}, we describe the DKIST observations and the two sets of simulations including the details of the spectral line synthesis. Section~\ref{sec:res} provides a detailed description of the events detected in the observations and comparison with similar events in the simulated data, aimed at a more detailed interpretation of the observations. In Section~\ref{sec:disc}, we discuss the results including the energy estimates of the detected events and their role in chromospheric heating.

\section{Observations and Simulations} \label{sec:data}
\subsection{DKIST observations with VBI} \label{subsec:DKIST_VBI}

On 2022 May 26 the DKIST observed a quiet Sun region at disk center between 17:46UT and 19:31UT (DKIST Observation ID: pid\_1\_36). The mean inferred Fried parameter during the extracted observing sequence was $11$~cm (at $500$~nm wavelength) reaching a value of $15$~cm, at its best period. Observations were carried out with the Visible Broadband Imager (VBI; \cite{2021SoPh..296..145W}), one of the first-light instruments at the DKIST. The VBI observes the Sun with high-speed 4k$ \times $4k cameras with a frame rate of $30$~s$^{-1}$. The pixel size at the recorded wavelengths corresponds to $\sim0.011$~arcsec ($8$~km) on the solar surface, corresponding to Nyquist sampling of DKIST’s theoretical diffraction limit in the G-band. The instrument acquired images alternating between two visible wavelength filters, with passbands for the G-band (CH\,molecular bandhead at around $430$\,nm) and the \ion{Ca}{2}\,K spectral line ($393$\,nm). The G-band filter has a passband width of $0.5$\,nm full-width at half-maximum (FWHM), while the \ion{Ca}{2}\,K filter has $0.1$\,nm FWHM. These filters provide information about the plasma of the deep photosphere and the upper photosphere/low chromosphere, respectively. The same data set (pid\_1\_36) was analyzed with respect to photospheric bright points by \citet{2026ApJ...999..201K}.

The data were dark and flat-field corrected prior to undergoing speckle reconstruction \citep{2008A&A...488..375W}. The speckle image reconstruction process uses 80 short exposed frames ($25$~ms exposure time for \ion{Ca}{2}\,K and $1$~ms for the G-band channel) as input to estimate the Fourier phases and amplitudes of the object using statistical analysis. The result of the processing is a single, high-resolution image calibrated with the best estimate of the system transfer function to achieve a high photometric accuracy in the reconstruction. The resulting cadence for the \ion{Ca}{2}\,K time series and the G-band time series after reconstruction is approximately $6$~s for both. 

The post-speckle reconstructed images were co-aligned prior to being destretched \citep{2010SoPh..261..363J} by determining local offsets between successive images. This corrects for any residual spatial distortions present as a result of, e.g., atmospheric turbulence, for the duration of the observations, which were not corrected in-situ at the telescope (e.g., through the use of adaptive optics). The final step involved co-aligning the G-band and \ion{Ca}{2}\,K images with respect to each other, using long-lived co-spatial features, to allow a comparison of features co-spatially between the filter images. 

A sequence with the best seeing, spanning a FoV of $12.76 \times 12.76$\,Mm$^{2}$ and with a duration of $26$~minutes, is selected for the analysis. The data are one of the few quiet Sun disk-center data sets at such a high cadence and with good seeing available so far from DKIST. Although data were simultaneously acquired with the Visible Spectropolarimeter \citep[ViSP;][]{2022SoPh..297...22D} instrument, our events did not fall within its FoV.  
\begin{figure*} [h!]
	\centering
	\includegraphics[width=\textwidth]{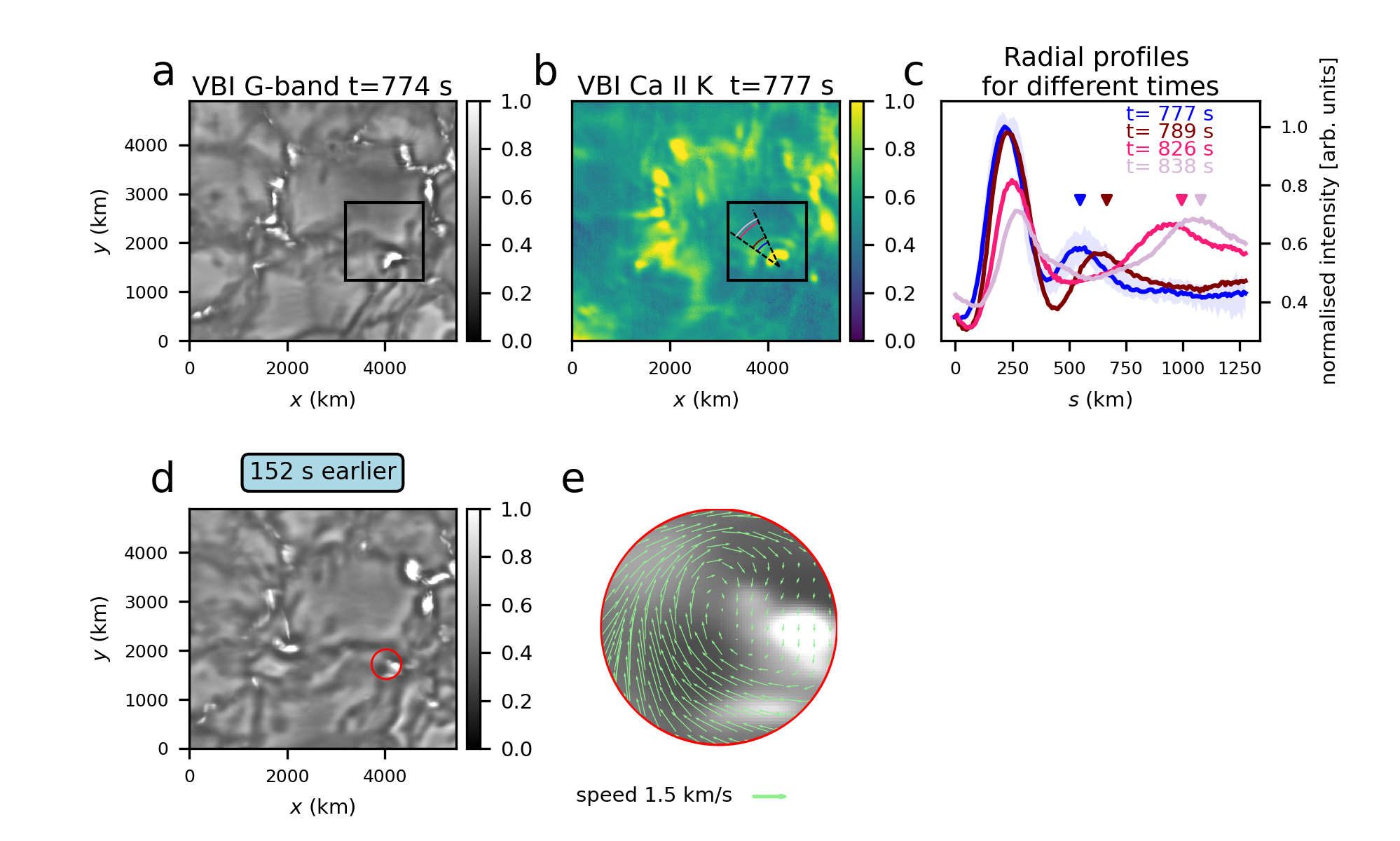}
	\caption{Wave front event Nr.~E1 with the G-band bright point in the photosphere, the associated \ion{Ca}{2}\,K bright grain in the low chromosphere, and the emitted arc-shaped wave front. (a) Filter image of the VBI G-band channel showing the deep photosphere. The central granule is surrounded at its periphery by bright point-like features, which are manifestations of small-scale magnetic flux concentrations. The bright point connected to the wave phenomena discussed here is located within the black square. (b) Filter image of the VBI \ion{Ca}{2}\,K channel showing the low chromosphere/upper photosphere. Within the black square, the bright chromospheric grain is partially surrounded by a bright, arc-shaped front traveling away from it. The coloured arcs indicate the position of the bright front at subsequent times (see panel~(c) for times). (c) Radiative intensity of the \ion{Ca}{2}\,K channel (normalised to the maximum intensity of the \ion{Ca}{2}\,K bright grain) along the radial direction of the travel path of the bright front. Each profile is the mean of profiles taken in 5{\textdegree} steps between the black dashed radial lines indicated in panel~(b). The shaded region around the blue curve ($t = 777$~s) displays the minimum and maximum of the intensity for all curves at that time step, shown as a representative example. The position of the bright front, as inferred from the applied Gaussian fit, is marked with triangles, whereby the colours denote the time step of the observation, indicated by labels. The peak intensity at $s \sim 250$~km corresponds to the bright \ion{Ca}{2}\,K grain from where the arc emanates. (d) G-band image recorded approximately $152$~s earlier than the G-band snapshot shown in panel~(b). (e) Close-up of the region of interest shown in panel~(d). Green arrows display the apparent horizontal velocities obtained through a Local Correlation Tracking (LCT) method (see Section~\ref{subsec:BB_imagery} for details). The velocities have been averaged over a time range of $60$~s centered at the time $t = 622$~s. An animation of the maps of panels (a) and (b) is linked to this figure. The animation loops three times and shows a $207$~s long sequence, for which the black rectangles and the dashed colored lines have been omitted. A black circle highlights the event location. A pink star indicates the location of the bright point at several selected frames of the \ion{Ca}{2}\,K maps and a pink dashed arc denotes the location of the bright front at times $t = 771$~s, $t = 813$~s, and $t = 844$~s.  }
	\label{fig:Obs_ex1_tdev}
\end{figure*}
\subsection{Simulations and synthesis} \label{subsec:sims}
Within this study, we employ two separate simulation data to compare with the DKIST observations. They were obtained using different initial magnetic field configurations and boundary conditions and generated by two fairly different, fully self-consistent, numerical radiation magnetohydrodynamic simulation tools, namely, CO5BOLD \citep{2012JCoPh.231..919F} and Bifrost \citep{2011A&A...531A.154G}. The CO5BOLD simulation snapshots used (ID: d3gt57g44v50fc\_high\_cad\_sec) have a duration of $15$~minutes with a cadence of $2$~s and with the computational domain spanning $9.6$~Mm $\times$ $9.6$~Mm $\times$ $2.8$~Mm. The computational cell size is $10$~km in each spatial direction. The vertical extent spans from $z \approx -1240$~km to the top boundary at $z \approx 1560$~km, where $z = 0$ corresponds to the mean Rosseland optical depth $\tau_{\rm R} = 1$. The initial state contained a homogeneous, vertical magnetic field of $50$\,G magnetic flux density, which was relaxed over $1.5$\,h of physical time until the beginning of the analysed time series. Continuum and \ion{Ca}{2}\,K maps
derived from these data
were synthesized using the Lightweaver \citep{2021ApJ...917...14O} code. The \ion{Ca}{2}\,K spectral line was synthesised over a wavelength range of $\pm 0.5$\,nm around the central wavelength of $393.3663$~nm. The synthesis was carried out assuming complete frequency redistribution (CRD)
but in certain cases, effects of partial frequency redistribution (PRD) are taken into account.  Electron densities are calculated under local thermodynamic equilibrium (LTE) conditions.
The \ion{Ca}{2}\,K maps are then constructed by applying the VBI transmission profile. The continuum 
maps are synthesised at a wavelength of $450.287$~nm. A detailed description of the CO5BOLD simulation tool is given in \cite{2012JCoPh.231..919F}.

Within this study we also use the publicly available Bifrost simulation data with identity number ch024031\_by200bz005. A detailed description of the Bifrost simulation tool is given in \cite{2011A&A...531A.154G}. The Bifrost data cube has a cadence of $10$~s and a range of $24$~Mm $\times$ 24~Mm $\times$ 17~Mm ($768 \times 768 \times 768$ grid points) with grid cells of $31.25$~km in the horizontal direction. The range in geometrical height is from $z \approx -2.5$~Mm to $z \approx 14.5$~Mm with respect to the solar surface (i.e., optical depth, $\tau = 1$ at $500$~nm). In this case, a magnetic field configuration with an average signed strength of $5$~G at $\tau_{500} = 1$ is used to represent a coronal hole. At the bottom boundary, a horizontal field of $200$~G is applied to the inflows. This setup allows the field strength to slowly increase over time, with interactions between the existing field and the new flux. For the present work we then use a time period of 400~s in which the average unsigned field strength at $\tau_{500} = 1$ is $40$~G  (frames $420$\,--\,$439$). For the synthesis of the spectral lines, the RH code \citep{2001ApJ...557..389U, 2015A&A...574A...3P} was employed. \\

In both cases, near the solar surface ($\tau_{\rm R} = 1$), initially imposed magnetic field is quickly advected to the intergranular lanes by the granular flow. There, the dynamic pressure concentrates it to dynamical equipartition field strength of several hundered Gauss. Further intensification to kG field strength proceeds by the convective instability in the presence of magnetic fields, a process known as convective collapse \citep{1979SoPh...61..363S}. Also in both cases, vortex flows and associated magnetic twists naturally appear and are not deliberately excited. They are produced by the convective dynamics as a result of the self-consistent numerical solution of the magneto-hydrodynamic (MHD) equations together with the equations of radiative transfer. More information on the simulations employed and on the differences between the two simulation data can be found in Appendix~\ref{append:diff_sim}.


\section{Results} \label{sec:res}

\subsection{Observations}

\subsubsection{Broadband imagery in the \ion{Ca}{2}\,K and G-band channels of VBI} \label{subsec:BB_imagery}
We detect propagating fronts of enhanced radiative intensity in the high-cadence \ion{Ca}{2}\,K sequence. Their distinct arc shapes emanate from their source location, where we find one or more steadily evolving chromospheric bright grains with corresponding bright points (BPs), in the G-band filter image.

In Figure~\ref{fig:Obs_ex1_tdev}, we present the observation of an example propagating front, denoted as event Nr.~E1.  
The observation times given throughout the text and in the figures and animations are the times given in the full sequence shown in the animation linked to Figure~\ref{fig:Obs_collage}.   

\begin{figure*} [h!]
	\centering
	\includegraphics[width=\textwidth]{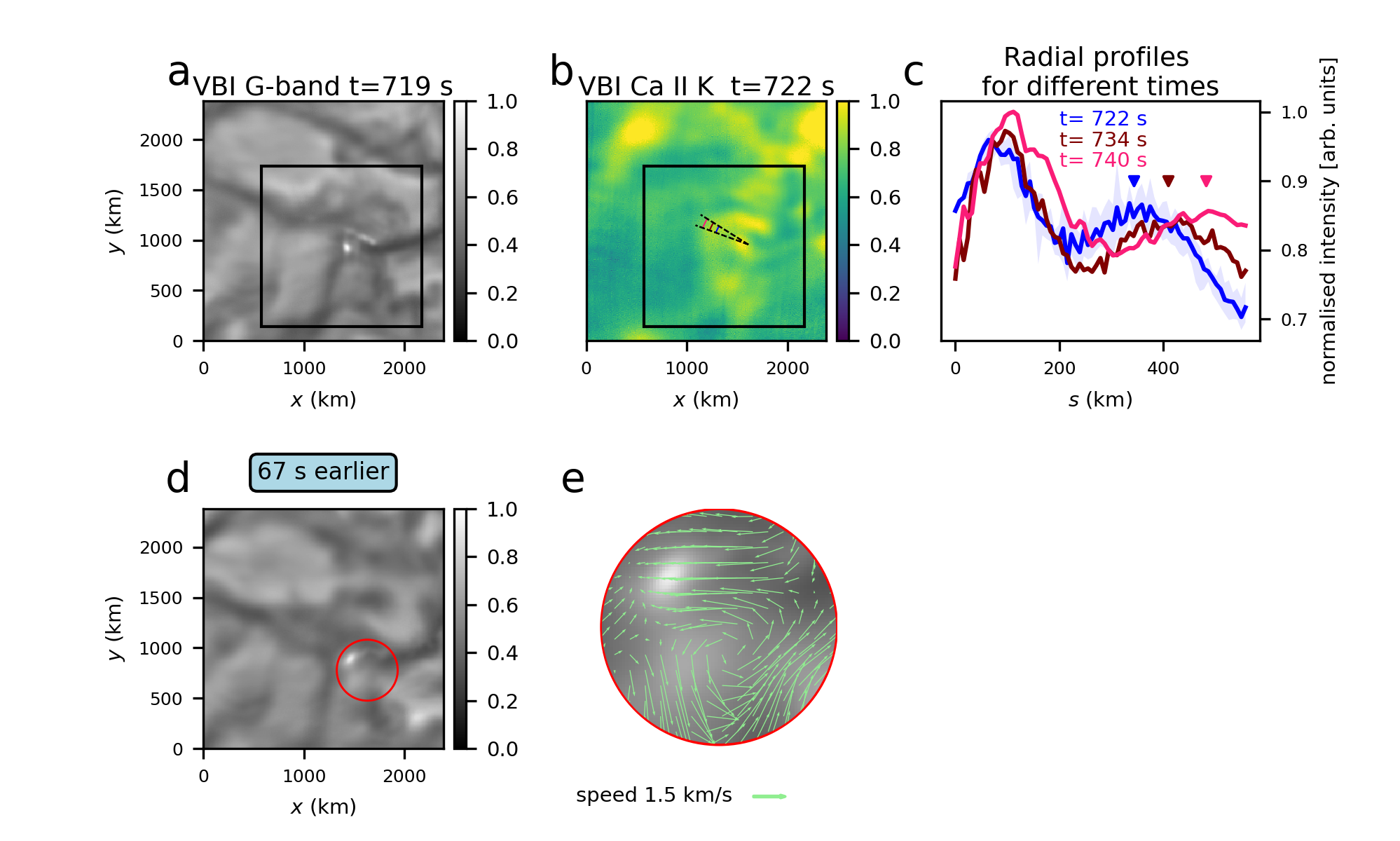}
	\caption{Event Nr.~E2 of the DKIST VBI examples showing the magnetic bright point located within the region of influence of a vortex in the photosphere, its corresponding signature in the low chromosphere, and the associated bright front signature. The figure is in the same format as Figure~\ref{fig:Obs_ex1_tdev}. For panel (e) the velocities have been averaged over a time range of $30$~s centred at the time $t = 646$~s. An animation of the images shown in panels (a) and (b) is available and linked to this Figure. The animation loops three times and shows a $140$~s long sequence, for which the black rectangles and the dashed colored lines have been omitted. A black circle highlights the event location. A pink star indicates the location of the bright point at several selected frames of the \ion{Ca}{2}\,K maps and a pink dashed arc denotes the location of the bright front at $t = 698$~s, $t = 716$~s, and $t = 734$~s. }
	\label{fig:Obs_ex2_tdev}
\end{figure*}

Figure~\ref{fig:Obs_ex1_tdev}(b) shows a snapshot of a bright arc, visible from the \ion{Ca}{2}\,K filter image series. The bright arc at time $t = 777$~s is delineated by a blue circular segment and its subsequent development in maroon, pink, and lavender corresponding to times $t = 789$~s, $t = 826$~s, and $t = 838$~s, respectively. The bright front starts from the \ion{Ca}{2}\,K bright grain located at $(x,y) = (4200, 1700)$~km at approximately time $t = 759$~s. The full time sequence of the representations shown in panels (a) and (b) can be followed with the animation linked to Figure~\ref{fig:Obs_ex1_tdev}, and an enlarged version of the \ion{Ca}{2}\,K image within the black square is repeated in the top left panel of Figure~\ref{fig:Obs_collage}.

The FoV of the G-band filter image shown in Figure~\ref{fig:Obs_ex1_tdev}(a) is dominated by a large granule surrounded by BPs residing in the adjacent intergranular lanes, which give rise to the bright grains seen in the \ion{Ca}{2}\,K filtergrams.
An automated feature identification and tracking algorithm \citep{2010ApJ...722L.188C} was utilized to isolate BPs within the G-band images. This algorithm isolates local intensity maxima within the intergranular lanes, followed by performing intensity cross-sections in different directions at 5{\textdegree} intervals about the barycenter location in order to accurately determine the BP boundary. Measurement of intensity gradients in these cross-sections enable the algorithm to discard false detections (e.g., fragmented granules). The tracking phase of the algorithm looks for similar features that persist between multiple frames in the vicinity of the originally detected BP. The location and size of these identified and tracked BPs are then stored for further analysis, where the sequence of locations serve to determine the velocities and paths of tracks of the BPs \citep{2011ApJ...740L..40K}. Features were identified and tracked in the G-band images to determine the relationship between the bright, propagating arcs observed in the \ion{Ca}{2}\,K images and the underlying magnetic field represented by the BPs. 
The BP shown in the black square of Figure~\ref{fig:Obs_ex1_tdev}(a) is the result of two BPs that merged at $t = 756$~s, after one of them experienced a rotating motion. The automated BP tracking reveals that this rotating BP first accelerates to a transverse velocity of $1.1 \pm 0.2$~km\,s$^{-1}$ before entering a $66$~s long rotational phase with an angular velocity of $1.1 \pm 0.1$~deg\,s$^{-1}$. The angular velocity is inferred from the motion of the BP over this time window using the barycentres identified from the BP tracking routine.

The FLCT (Fourier local correlation tracking) code \citep{2008ASPC..383..373F} was applied to the G-band time series to obtain apparent horizontal velocities. The FWHM of the gaussian employed for weighing of the correlated sub-images was set to $\sigma = 20$~pixels. For event Nr.~E1 (Figure~\ref{fig:Obs_ex1_tdev}(e)), we averaged the obtained velocities over a time range of $61$~s over the rotation phase of the BP. The output of the FLCT code indicates a vortical plasma motion (swirl) during the same time period. This can be seen by visual inspection of the horizontal velocity field that displays a coherent circulating pattern of velocity vectors co-located with the BP (see Figure~\ref{fig:Obs_ex1_tdev}(e)). This  rotation of the surrounding plasma takes place approximately two minutes prior to the launch of the wave in the chromosphere. The swirl is likely the underlying cause of the rotational motion of the BP, as is often seen for magnetic flux concentrations due to the ubiquitous presence of small-scale swirling motions within quiet Sun intergranular lanes \citep{2008ApJ...687L.131B,2019A&A...632A..97L,2023SSRv..219....1T}.

At close inspection, the bright arc in the \ion{Ca}{2}\,K filtergrams starts as a small, slightly curved brightening at $t = 759$~s on the left side adjacent to the BP that had a rotational component to its motion, before traveling as a semi-circular arc across the large granule. We find a mean horizontal propagation speed 
of the bright front
of $ 9.2 \pm 4.7$~km\,s$^{-1}$  (see Table~\ref{tab:results} for inferred parameters and associated deviation limits). 

The intensity profile of the bright front along its direction of motion can be approximated with a Gaussian curve. The FWHM of this intensity profile is initially around $200$~km, and expands in the last time steps to a mean width of $430$~km (Figure~\ref{fig:Obs_ex1_tdev}(c)). We clarify here in advance and will show further below that simulations suggest this to be the intensity distribution over a shock-front surface and therefore should not be mistaken with the actual shock structure. It displays a fine structure within the arc with a striation consisting of bright and dark ridges, best visible from time $t=807$\,s onwards, which may point to the onset of a corrugation instability of the shock surface. The analysis of such an instability and also the possible connection with the magnetic structure of the environment is left for future investigations with the availability of spectrapolarimetric data. At $t = 856$~s, the arc has traveled to a distance corresponding to approximately the middle of the underlying photospheric granule. At this time, several brightenings can be distinguished within and around the arc before it diffuses and becomes untraceable. 

Table~\ref{tab:results} lists the apparent horizontal propagation speed, FWHM, and radial distance of the bright front relative to the source location for observation and simulation events. We select seven radial directions from the chromospheric bright grain to the arc-shaped bright front, separated by 5{\textdegree} intervals for observations event Nr.~E1 for determining intensity profiles. For events Nr.~E2 and E3, we also choose seven radial directions with intervals of 1{\textdegree}  with the exception of the first and last interval of event Nr.~E2 with 2{\textdegree} and 12{\textdegree}, respectively. The radial intensity profiles are fitted using the lmfit python routines and employing a Gaussian fit to the radial profile of the bright front. The front position (maximum of the Gaussian fit) and the FWHM are determined by fitting the mean of the seven profiles. The errors are the standard deviation of the fitting results obtained from the individual intensity profiles. We compute the speed of the bright front from the maxima found with the Gaussian fit to the mean profile for the time steps listed in Table~1. The average speed is the mean of these speeds whereby the error is the standard deviation of the individual speeds obtained for the mean profiles. The same method was applied to the simulation data presented in Sect.~\ref{subsec:Numerical_sims} with the radial profiles taken at 1{\textdegree} intervals for simulation event Nr.~S1 as well as for simulation event Nr.~S2.

\begin{table}
    \centering
    \begin{tabular}{cccc}
\hline
      {Event Times}  &  {Speed (km\,s$^{-1}$)} & {FWHM (km)} & {Maximum Position (km)}\\
\hline\hline 
      {\textit{Obs. Event Nr.~E1} }   & $9.2\pm4.7$ &  & \\
        $t = 777$~s & & $215.9 \pm 35.3$ & $548.3 \pm 17.8$ \\
        $t = 789$~s & & $258.1 \pm 72.2$${^*}$ & $673.1 \pm 45.9$${^*}$ \\
        $t = 826$~s & & $406.4 \pm 82.2$ & $950.5 \pm 35.7$ \\
        $t = 838$~s & & $429.5 \pm 66.5$ & $1068.1 \pm 43.6$\\
         
    \hline
      {\textit{Obs. Event Nr.~E2} }   & $7.5\pm4.4$ &  & \\
       $t = 722$~s  & & $266.9 \pm 10.2$ & $344.4 \pm 11.1$ \\
       $t = 734$~s  & & $279.1 \pm 10.2$ & $405.8 \pm 18.1$ \\
       $t = 740$~s  & & $289.7 \pm 9.7$ & $465.4 \pm 19.8$ \\
    \hline
      {\textit{Obs. Event Nr.~E3} }   & $6.3\pm1.1$ &  & \\
       $t = 466$~s  & & $182.7 \pm 10.1$ & $280.2 \pm 2.6$ \\
       $t = 472$~s  & & $185.7 \pm 5.0$ & $325.8 \pm 3.2$ \\
       $t = 479$~s  & & $186.1 \pm 14.5$ & $340.0 \pm 4.1$ \\
       $t = 485$~s  & & $103.4 \pm 96.4$ & $396.4 \pm 3.0$ \\
    \hline
      {\textit{Sim. Event Nr.~S1}}   & $7.6\pm0.8$ &  & \\
       {\textit{(CO5BOLD)}}  &  &  & \\
       $t = 5764$~s  & & $156.6 \pm 4.8$ & $372.5 \pm 3.3$ \\
       $t = 5770$~s  & & $182.4 \pm 2.0$ & $361.9 \pm 1.4$ \\
       $t = 5776$~s  & & $193.1 \pm 7.3$ & $406.5 \pm 3.0$ \\
       $t = 5782$~s  & & $212.8 \pm 4.4$ & $464.7 \pm 2.8$ \\
    \hline  
      {\textit{Sim. Event Nr.~S2}}   & $15.2\pm1.1$ &  & \\
       {\textit{(Bifrost)}}  &  &  & \\
       $t = 4280$~s  & & $186.7 \pm 2.8$ & $343.8 \pm 4.7$ \\
       $t = 4290$~s  & & $136.4 \pm 4.9$ & $477.1 \pm 4.0$ \\
       $t = 4310$~s  & & $118.2 \pm 8.2$${^*}$ & $798.0 \pm 13.0$${^*}$ \\
       $t = 4320$~s  & & $191.7 \pm 8.1$ & $960.6 \pm 14.9$ \\       
\hline
\multicolumn{4}{l}{\small ${^*}$6 profiles used for the error calculation.} \\
    \end{tabular}
    \caption{Apparent horizontal propagation speed, FWHM, and radial distance of the bright front relative to the source location for observation and simulation events. The times refer to the times of the \ion{Ca}{2}\,K frames within the full sequence animation (animation linked to Figure~\ref{fig:Obs_collage}).}
    \label{tab:results}
\end{table}



\subsubsection{Additional similar events in the observation data}
Nine more events exhibiting an evolution similar to event Nr.~E1 are detected within the examined image sequence. This implies a rate of $0.14$ events per Mm$^2$ per hour within the sequence. These examples are not as prominent as the example seen in event Nr.~E1 and decay faster. Sample images for all $10$ events are shown in the Appendix~\ref{append:additional_examples} as Figure~\ref{fig:Obs_collage}, while results for the apparent horizontal propagation speed, FWHM, and position of the wavefront are provided in Table~\ref{tab:results} for event Nrs.~E1-E3 for which these quantities are best identifiable.

\begin{figure*} [h!]
	\centering
	\includegraphics[width=\textwidth]{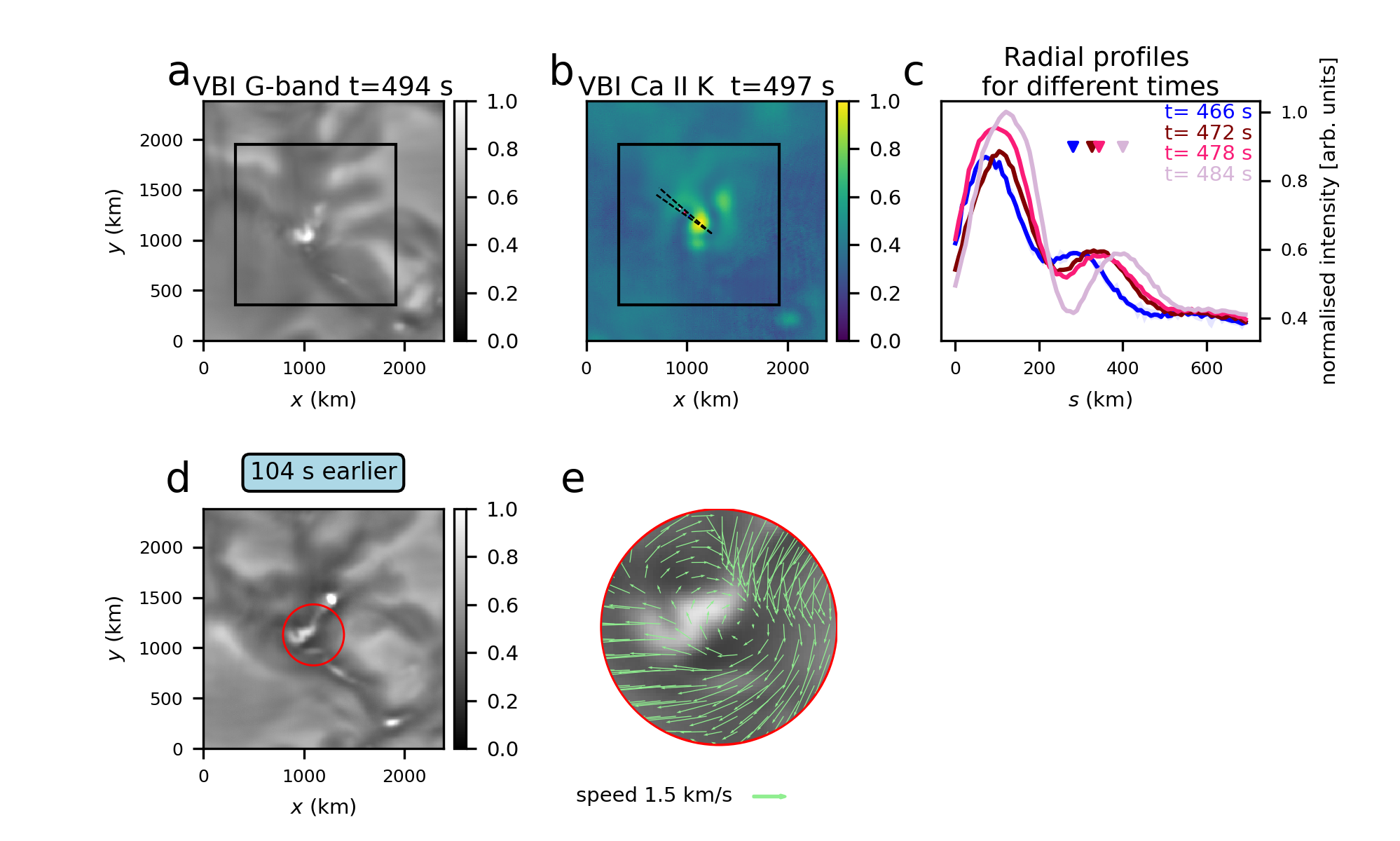}
	\caption{Event Nr.~E3 of the DKIST VBI examples showing the magnetic bright point located within the region of influence of a vortex in the photosphere, its corresponding signature in the low chromosphere, and the associated bright front signature. The figure is in the same format as Figure~\ref{fig:Obs_ex1_tdev}. For panel (e) the velocities are shown for the time $t = 390$~s. An animation of the images shown in panels (a) and (b) is available and linked to this Figure.  The animation loops three times and shows a $122$~s long sequence, for which the black rectangles and the dashed colored lines have been omitted. A black circle highlights the event location. A pink star indicates the location of the bright point at several selected frames of the \ion{Ca}{2}\,K maps and a pink dashed arc denotes the location of the bright front at $t = 503$~s, $t = 521$~s, and $t = 539$~s.}
	\label{fig:Obs_ex3_tdev}
\end{figure*}
Events Nr.~E2 and E3 are presented in the same manner as in Figure~\ref{fig:Obs_ex1_tdev}, including their animation in Figure~\ref{fig:Obs_ex2_tdev} and Figure~\ref{fig:Obs_ex3_tdev}, repectivly. They also show in the \ion{Ca}{2}\,K filter images an arc-shaped enhanced wave front originating from a \ion{Ca}{2}\,K bright grain that coincides with one or several photospheric BPs. Figure~\ref{fig:Obs_ex2_tdev} (event Nr.~E2) shows the time instance of an event in which a chain of BPs trace the advecting swirling motion (see G-band images of the full time sequence in the animation linked to Figure~\ref{fig:Obs_ex2_tdev} and the flow field obtained from local correlation tracking in Figure~\ref{fig:Obs_ex2_tdev}(e)). The \ion{Ca}{2}\,K bright front launched at approximately $t = 691$~s, a minute after the appearance of the swirl, is much fainter than in event Nr.~E1, yet, clearly visible. We are able to trace the front, of width $266.8 \pm 10.2$~km, for about $50$~s in which time span it propagates at an apparent horizontal speed of $7.5 \pm 4.4$~km\,s$^{-1}$ before becoming less discernible when merging with surrounding brightenings. Event Nr.~E3 shows a BP caught in a vortex and assuming a ‘U’-shaped form (see Figure~\ref{fig:Obs_ex3_tdev} and accompanying animation). The evolution of all $10$ events is shown in the animation of Figure~\ref{fig:Obs_collage}. Although the events exhibit a similar overall behaviour, we see a variety in visibility, life-time, propagation speed, and even shape of the bright front with some being close to circular, and others becoming visible in different locations along the arc as time progresses. We will return to this observation in Section~\ref{subsec:Topography}, where we speculate on the possible cause of these differences.

For events Nr.~E1 and E3 (Figure~\ref{fig:Obs_ex1_tdev} and Figure~\ref{fig:Obs_ex3_tdev}), the BP is close to the swirl center — this configuration very likely excites a torsional Alfv{\'e}n pulse. The time delay between the appearance of the photospheric swirl to the launch of the bright arc seen in \ion{Ca}{2}\,K is $\sim130$~s for event Nr.~E1 and $\sim70$~s for event Nr.~E3. For event Nr.~E2 the situation is different. Rather we find there a configuration where a chain of small G-band BPs move relative to a single BP both being advected by the velocity field in the periphery of a vortex. That relative motion is likely sufficient to excite a torsional wave or pulse but we cannot exclude that other modes, such as sausage, kink and/or flute are excited as well. In fact, such relative motion of G-band bright points have been observed to be the photospheric basis of chromospheric swirls \citep{2009A&A...507L...9W}, which are typically a signature of torsional Alfv{\'e}n pulses \citep{2021A&A...649A.121B}. For observation case Nr.~E2, the time delay between the appearance of the photospheric swirl to the launch of the bright arc seen in \ion{Ca}{2}\,K is $\sim45$~s. We observe in $7$ out of $10$ cases a vortical motion in the photosphere prior to the emission of the bright front — in three cases a swirl is not easily identified. The temporal distance between the occurrence of the vortex and the bright front is estimated by the time difference between the full development of the swirl in the horizontal velocity field as derived with correlation tracking from the G-band image sequence and the first signature of the bright front as detected by visual inspection of the \ion{Ca}{2}\,K image sequence. The time delay for all cases is between about 1 and 2 min, which approximately corresponds to the distance between photosphere and low chromosphere when adopting a reasonable characteristic speed of $10$~km\,s$^{-1}$.
Therefore, we consider the connection between vortical flow and launch of the bright front as very probable, although we cannot prove it from the observational data alone.

\begin{figure*} [h!]
	\centering
	\includegraphics[width=\textwidth]{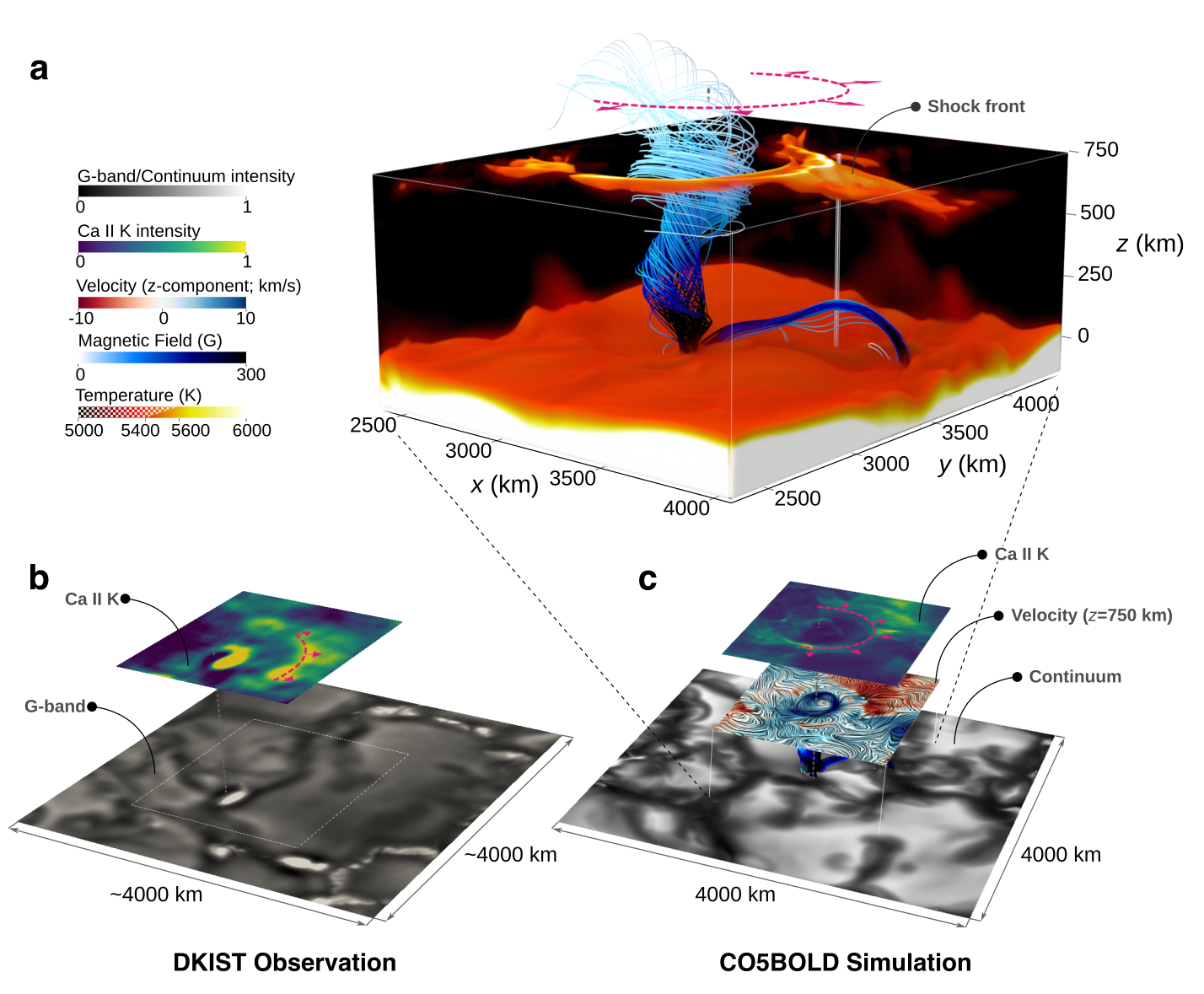}
	\caption{Comparison of the observational case Nr.~E1 with a simulated atmosphere and synthesized blue continuum and \ion{Ca}{2}\,K maps. (a) Volume rendering of the temperature above $5000$~K of simulation event Nr.~S1 (at $t = 5814$~s) with blue magnetic field lines representing the twisted magnetic vortex structure. The bright front appears as a thin, arc-shaped temperature enhancement on one side of the vortex illustrated by the dashed pink semi-circle with arrows indicating the direction of propagation. The shock front proper is visible as the sheet of enhanced temperature slightly inclined with respect to a horizontal plane. The location of the line-of-sight used in Figures~\ref{fig:Sim_time_distance_temp} and \ref{fig:Shockfront} at $(x,y) = (3435~\mathrm{km}, 3755~\mathrm{km})$ is shown as a white line on the right hand side in the 3D view. (b) DKIST/VBI observations of event Nr.~E1 at time $t = 826$~s. The pink dashed arc shown in the \ion{Ca}{2}\,K image represents the location of the bright front, while arrowheads indicate the direction of travel of the front. (c) The top map shows the line-core \ion{Ca}{2}\,K map referring to the low chromosphere/upper photosphere and the bottom map shows the continuum intensity at $450$~nm referring to the nearly same photospheric region as the observed G-band image, both of which were computed from the numerical simulation data. The middle map displays the line integral convolution of the horizontal velocity field at $z = 750$~km with the colours denoting the vertical velocity, $v_z$, where red indicates downflows and blue indicates upflows. An animation of panel (a) is linked to this Figure. The time sequence spans a duration of $68$~s and shows the evolution of the temperature and magnetic lines of force. In the time sequence the bright front appears as a thin, arc-shaped temperature enhancement on one side of the vortex at the top boundary of the subdomain. The shock front appears as an extended sheet of enhanced temperature, slightly inclined with respect to the horizontal plane.}
	\label{fig:Sim_ex1_overview_comp}
\end{figure*}
\subsection{Simulations}
\subsubsection{Bright front in numerical simulations} \label{subsec:Numerical_sims}

\begin{figure*} [h!]
	\centering
	\includegraphics[width=0.95\textwidth]{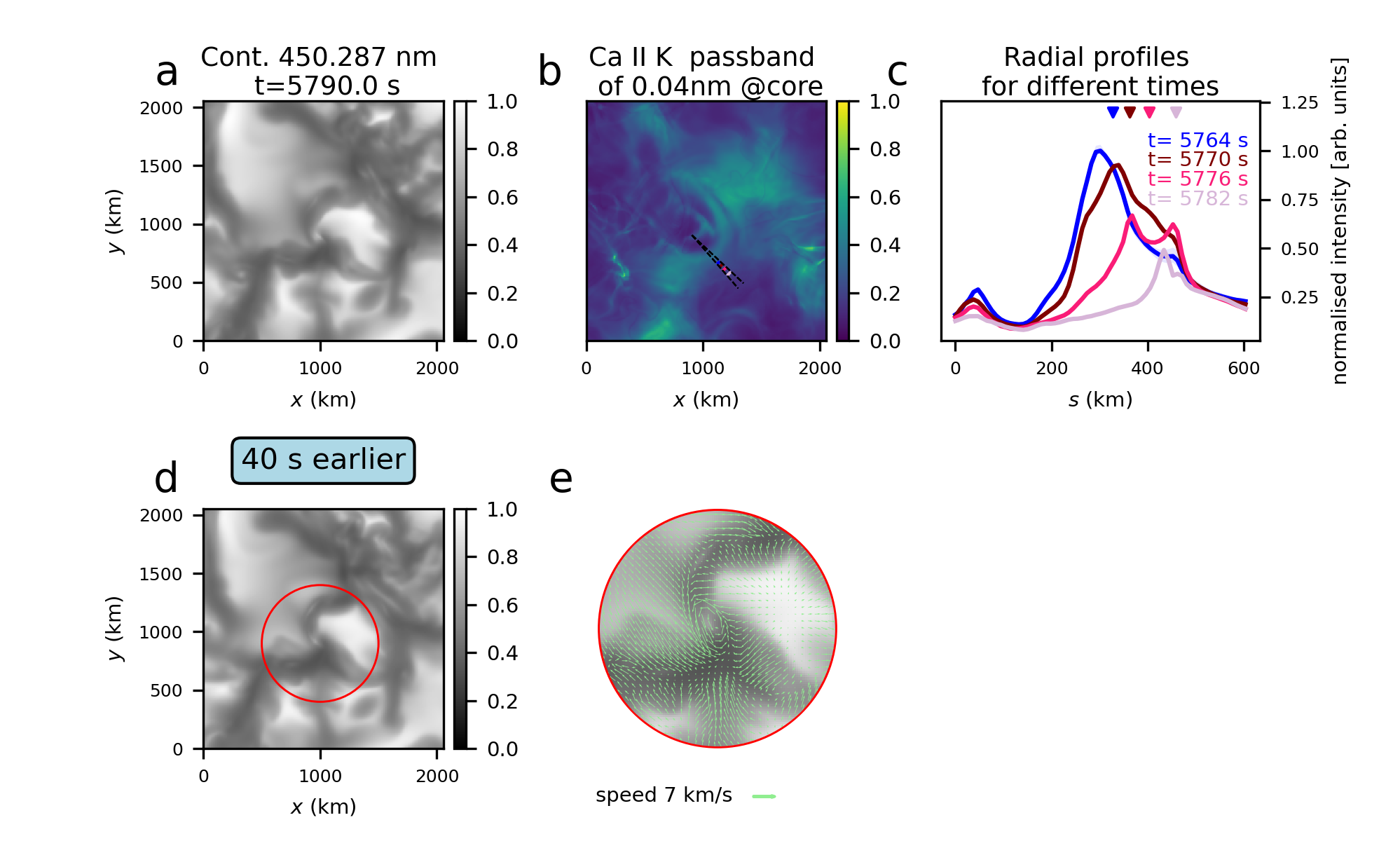}
	\caption{Simulation event Nr.~S1 in the same format as Figure~\ref{fig:Obs_ex1_tdev}. (a) Synthesized continuum map at the wavelength $450.287$~nm. The photospheric bright continuum point at $(x,y) = (1000~\mathrm{km},1000~\mathrm{km})$ is located at the source of the chromospheric arc-shaped front seen in panel (b). (b) Synthetic \ion{Ca}{2}\,K map computed under PRD. The coloured arcs indicate the position of the bright front at subsequent times (see panel (c) for times). (c) Normalized intensity of the \ion{Ca}{2}\,K map in panel (b) along the radial direction of the travel path of the bright front. Each profile is the mean of profiles taken in $1$ degree steps between the black dashed radial lines indicated in panel (b). The position of the bright front is marked with triangles, whereby the colours denote the time step indicated by labels. The procedure to obtain these values is equivalent to the procedure used for the observational analysis. (d) Continuum image computed $40$~s earlier than the \ion{Ca}{2}\,K snapshot shown in panel (b). (e) Close-up of the region of interest shown in panel (d). Green arrows display the horizontal velocities extracted directly from the simulation at height $z = 0$~km. For panel (e) the velocities are shown for the time $t = 5750$~s.
    An animation of the continuum and \ion{Ca}{2}\,K maps shown in panels (a) and (b) is available and linked to this Figure. In the time sequence of duration $82$~s, the bright arc-shaped front is marked in the \ion{Ca}{2}\,K map at times $t = 5800$~s, $t = 5808$~s, and $t = 5818$~s by a dashed pink arc in the replay. The pink star indicates the location of the photospheric bright point during the time span of the event.}
	\label{fig:Sim_ex1_tdev_vel}
\end{figure*}

\begin{figure*} [h!]
	\centering
	\includegraphics[width=0.95\textwidth]{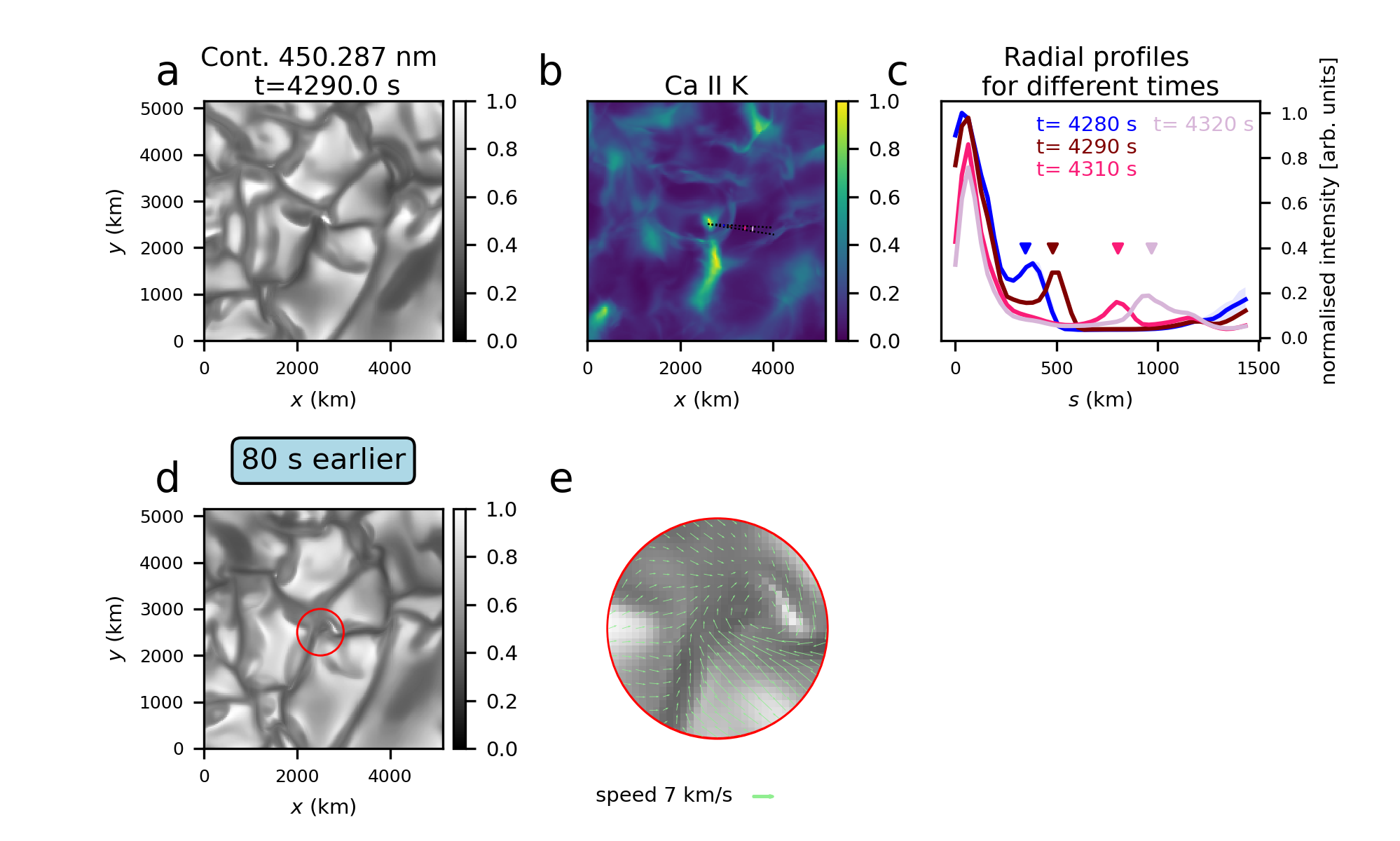}
	\caption{Simulation event Nr.~S2 in the same format as Figure~\ref{fig:Obs_ex1_tdev}. (a) Synthesized continuum map at the wavelength $450.287$~nm. The photospheric bright continuum point at $(x,y) = (2300~\mathrm{km},2600~\mathrm{km})$ is located at the source of the chromospheric arc-shaped front seen in panel (b). (b) Synthetic \ion{Ca}{2}\,K map computed under CRD. The coloured arcs indicate the position of the bright front at subsequent times (see panel (c) for times). (c) Normalized intensity of the \ion{Ca}{2}\,K map in panel (b) along the radial direction of the travel path of the bright front. Each profile is the mean of profiles taken in $1$ degree steps between the black dashed radial lines indicated in panel (b). The position of the bright front is marked with triangles, whereby the colours denote the time step indicated by labels. The procedure to obtain these values is equivalent to the procedure used for the observational analysis. (d) Continuum image computed $80$~s earlier than the \ion{Ca}{2}\,K snapshot shown in panel (b). (e) Close-up of the region of interest shown in panel (d). Green arrows display the horizontal velocities extracted directly from the simulation at height $z = 0$~km. For panel (e) the velocities are shown for the time $t = 4210$~s.
    An animation of the continuum and \ion{Ca}{2}\,K maps shown in panels (a) and (b) is available and linked to this Figure. In the time sequence of duration $390$~s, the bright arc-shaped front is marked in the \ion{Ca}{2}\,K map at times $t = 4290$~s and $t = 4300$~s by a dashed pink arc in the replay. The pink star indicates the location of the photospheric bright point during the time span of the event.}
	\label{fig:Sim_ex2_tdev_vel}
\end{figure*}

In the following, we concentrate on simulation event Nr.~S1 from the CO5BOLD data and simulation event Nr.~S2 from the Bifrost data. For simulation event S1, synthetic \ion{Ca}{2}\,K maps have been generated by applying the Lightweaver synthesis code \citep{2021ApJ...917...14O} to the full time series. Figure~\ref{fig:Sim_ex1_overview_comp}(c) shows a map of the \ion{Ca}{2}\,K intensity in the line core as computed under partial redistribution (PRD), while the right panel of the animation linked to Figure~\ref{fig:Sim_ex1_tdev_vel} shows the sequence of maps of the \ion{Ca}{2}\,K line computed under complete redistribution (CRD) and convolved with the VBI transmission profile using the approximation given in \cite{2014SPIE.9147E..9IW}.  
The \ion{Ca}{2}\,K maps of simulation event S2 were produced with the RH code \citep{2001ApJ...557..389U,2015A&A...574A...3P} under CRD, again applying the VBI filter profile and are shown in the right panel of the animation linked to Figure~\ref{fig:Sim_ex2_tdev_vel}. Both simulation sequences show wave signatures remarkably similar to those found in the observations, in particular, arcs of enhanced \ion{Ca}{2}\,K emission originating from magnetic flux concentrations and showing an apparent propagation in the horizontal direction. We thus encounter the same phenomenon whilst utilizing simulation data from two different simulation tools with different initial magnetic configuration and boundary conditions, which testifies to the robustness of the process. The continuum maps of simulation event Nr.~S1 reveal a bright point at the source location of the front. We also produced synthetic maps of the detailed G-band channel, which look very similar to the maps of the continuum at $450.287$~nm as expected. In particular, the continuum bright point is equally well visible as a G-band bright point. Appendix~\ref{append:snthetic_imagery} gives the synthetic imagery of an additional simulation event, Nr.~S3, found in the simulation data.

Figure~\ref{fig:Sim_ex1_tdev_vel} shows the bright front and its radial profiles in the same format as the observations. A photospheric swirl is located co-spatial with the magnetic flux concentration as seen in Figure~\ref{fig:Sim_ex1_tdev_vel} (d) and (e). The mean velocity of the front, as derived from the synthetic \ion{Ca}{2}\,K images is $7.6 \pm 0.8$~km\,s$^{-1}$. The width of the front increases from $156.6 \pm 4.8$~km to $212.8 \pm 4.4$~km in $18$~s. Simulation event Nr.~S2 also shows a continuum bright point at the source of its traveling bright front (see Figure~\ref{fig:Sim_ex2_tdev_vel}). The front exhibits a faster apparent horizontal propagation of $15.2 \pm 1.1$~km\,s$^{-1}$ with the width of the bright front changing only marginally during the sequence. These numbers are comparable to the corresponding numbers seen in observations within their range in variation.

\subsubsection{Physical topography of the bright front} \label{subsec:Topography}
The simulation data offer the possibility to study the bright front in more detail. Simulation event Nr.~S1 is showcased in Figure~\ref{fig:Sim_ex1_overview_comp}. The BP at the source location of the wave (Figure~\ref{fig:Sim_ex1_tdev_vel} and animation) exhibits a magnetic field strength of around $0.05$~T (see also the animation linked to Figure~\ref{fig:Sim_ex1_overview_comp}). Tracing the magnetic field lines (blue lines in Figure~\ref{fig:Sim_ex1_overview_comp}(a) reveals that they are twisted, implying the presence of a magnetic vortex \citep{2021ApJ...915...24S}. Examining the horizontal velocities at different heights of the model, we find that the twisted magnetic flux tube is located within a plasma swirl, present throughout the photosphere, and reaching the chromosphere. The swirl can be found in the middle image in Figure~\ref{fig:Sim_ex1_overview_comp}(c) that shows the line integral convolution of the horizontal velocity field for a specific height in the model atmosphere ($750$~km) superimposed with the vertical velocity, $v_z$, with blue colours denoting flows in the upward direction and areas in red denoting flows in the downward direction. From the animation linked to Figure~\ref{fig:Sim_ex1_overview_comp} one recognizes a temperature enhanced wave front (labelled ``Shock front” in Figure~\ref{fig:Sim_ex1_overview_comp}(a)) in the surrounding, ambient medium of the twisted magnetic flux concentration. The front is slightly inclined with respect to a horizontal plane, moving upward, and steepening into a hot, sharp front. Thus, from the simulation we conclude that it is a vertically propagating, inclined shock front that causes the observed apparent horizontally moving bright front, which in fact is the shock-front surface seen face-on. The vertical plasma motions around the shock front show downward flows upstream of the shock front and upward flows behind it. At this point, we can not compare this finding with observations for lack of information from spectral data but shall become a topic of future investigations.

\begin{figure*} [h!]
	\centering
	\includegraphics[width=\textwidth]{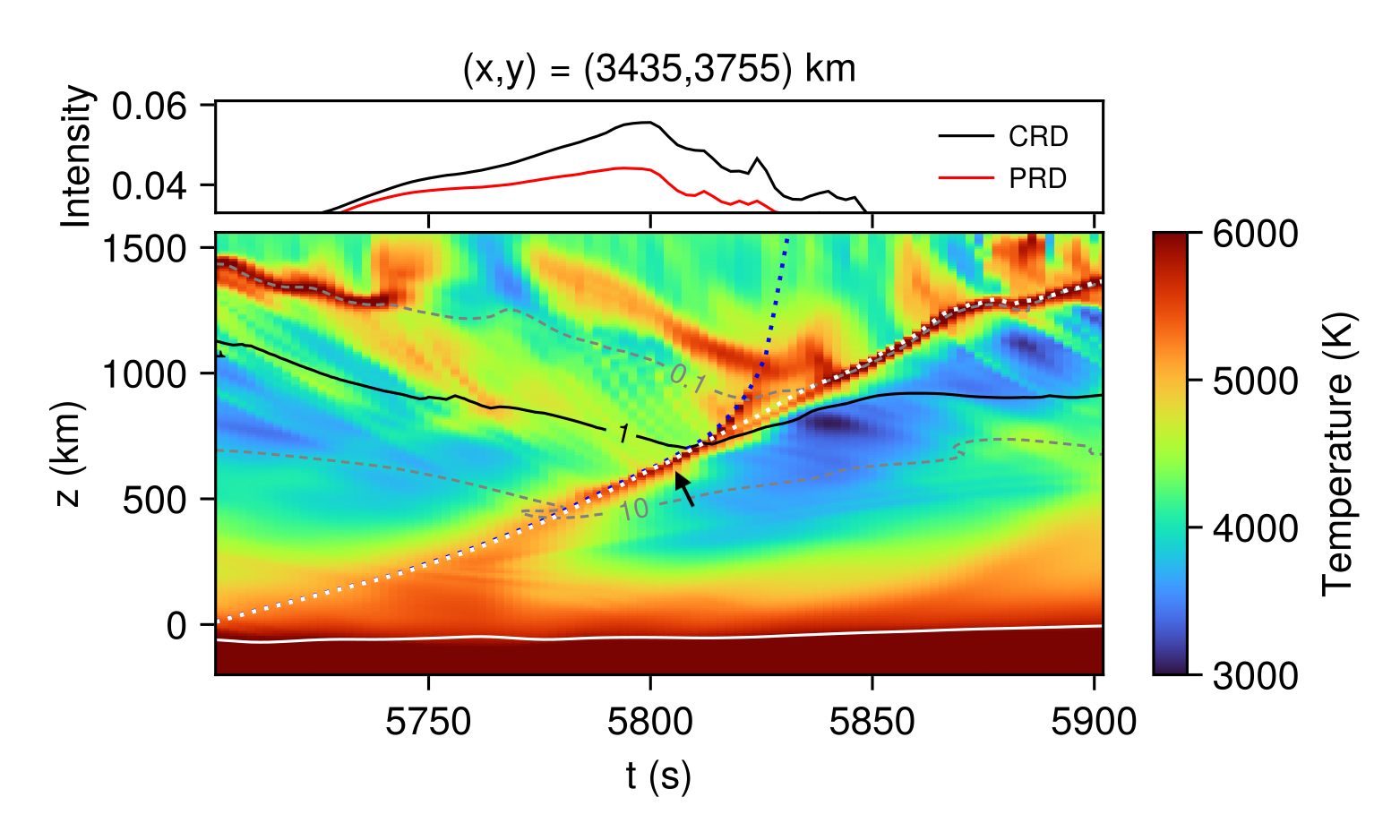}
	\caption{Time-distance diagram of the temperature from simulation event Nr.~S1. The diagram shows the passage of the shock front and the corresponding variation in \ion{Ca}{2}\,K filter intensity at location $(x,y) = (3435~\mathrm{km}, 3755~\mathrm{km})$ indicated by the vertical white line in Figure~\ref{fig:Sim_ex1_overview_comp}(a). The top panel shows the variation in the filter intensity (normalized to the intensity integrated over the full line width) calculated assuming complete frequency redistribution (CRD; black curve) and when taking partial frequency redistribution (PRD; red curve) into account. The bottom panel shows the temperature. The black arrow points at the temperature enhanced traveling shock front, which splits into two branches when crossing the $\beta = 1$ surface (black curve). The gray dashed contours indicate the layers of $\beta=0.1$ and 10. The white line indicates the Rosseland optical depth $\tau_{\rm R} = 1$. The dotted paths show the trajectory of a test particle moving with the fast speed $\sqrt{c_{\rm s}^{2}+\nu_{\rm a}^2}$ plus the line-of-sight component of the velocity (blue) and the sound speed ($c_{\rm s}$) plus the line-of-sight component of the velocity (white), while $\nu_{\rm a}$ is the Alfv{\'{e}}n speed. Up to the branching, the two trajectories stay very close to each other because the Alfv{\'{e}}n speed is low and therefore the wave being a fast, predominantly acoustic one.}
	\label{fig:Sim_time_distance_temp}
\end{figure*}

Figure~\ref{fig:Sim_time_distance_temp} shows the temperature and \ion{Ca}{2}\,K intensity as a function of time  along the line-of-sight at pixel location $(x, y) = (3435~\mathrm{km}, 3755~\mathrm{km})$, indicated by the white vertical line in Figure~\ref{fig:Sim_ex1_overview_comp}(a). There, we identify the shock wave as a fast magnetosonic, predominantly acoustic wave in a high plasma-$\beta$ ($\beta > 1$) environment (black arrow in Figure~\ref{fig:Sim_time_distance_temp}), where plasma-$\beta$ is the ratio of gas pressure to magnetic pressure. 
The plasma-$\beta$ is approximately equal to the ratio of sound speed to Alfv\'{e}n speed ($c_{\rm s}/v_{\rm A}$). The contours of $\beta=10$ and $\beta=0.1$ in Fig.\,\ref{fig:Sim_time_distance_temp} indicate the region where the plasma is high- and low-$\beta$, respectively. In the case of the high-$\beta$ regime (i.e., $\beta >$ 1, corresponding to $c_{\rm s}/v_{\rm A} > 1$), the sound speed is greater than the Alfv\'{e}n speed and therefore, the wave that follows the white dotted trajectory is mainly driven by gas pressure perturbation running at the fast (acoustic) speed and is therefore considered to be a fast, predominantly acoustic wave. We clearly see the temperature enhancement associated with this fast (predominantly acoustic) wave in the high-$\beta$ region. As this wave crosses the $\beta=1$ surface into the low-$\beta$ regime where the Alfv\'{e}n speed is larger than the sound speed, the wave splits into two branches: a fast wave, predominantly magnetic in nature, follows the blue dotted trajectory and a slow wave, predominantly acoustic in nature, closely follows the white dotted sound speed trajectory. More details on the shock wave identification are given in Appendix~\ref{append:energy_estimate} and corresponding Figures~\ref{fig:Sim_press_volume} and \ref{fig:Shockfront}.
The top panel in Figure~\ref{fig:Sim_time_distance_temp} captures the change in the \ion{Ca}{2}\,K filter intensity at the corresponding location as the shock front traverses the line-forming region. 

As the inclined shock front moves into the height range of the \ion{Ca}{2}\,K line formation, it closely mimics the observation by forming an arc-shaped front traveling in the horizontal direction as a consequence of projection effects. This also explains why the radiative intensity of the bright front may increase with time (as it does in observation case Nr.~E1) and not monotonically decrease as one might expect from a dissipating shock front. 
Both observation and simulation events show a variation in the width and appearance of the bright fronts seen in the \ion{Ca}{2}\,K diagnostics (see also Appendix~\ref{append:snthetic_imagery}). The interplay between shock propagation direction, projection effect, and spectral response is likely responsible for this variety.

\subsubsection{Torsional Alfv{\'e}n pulse} \label{subsec:Torsional}
\begin{figure*} [h!]
	\centering
	\includegraphics[width=\textwidth]{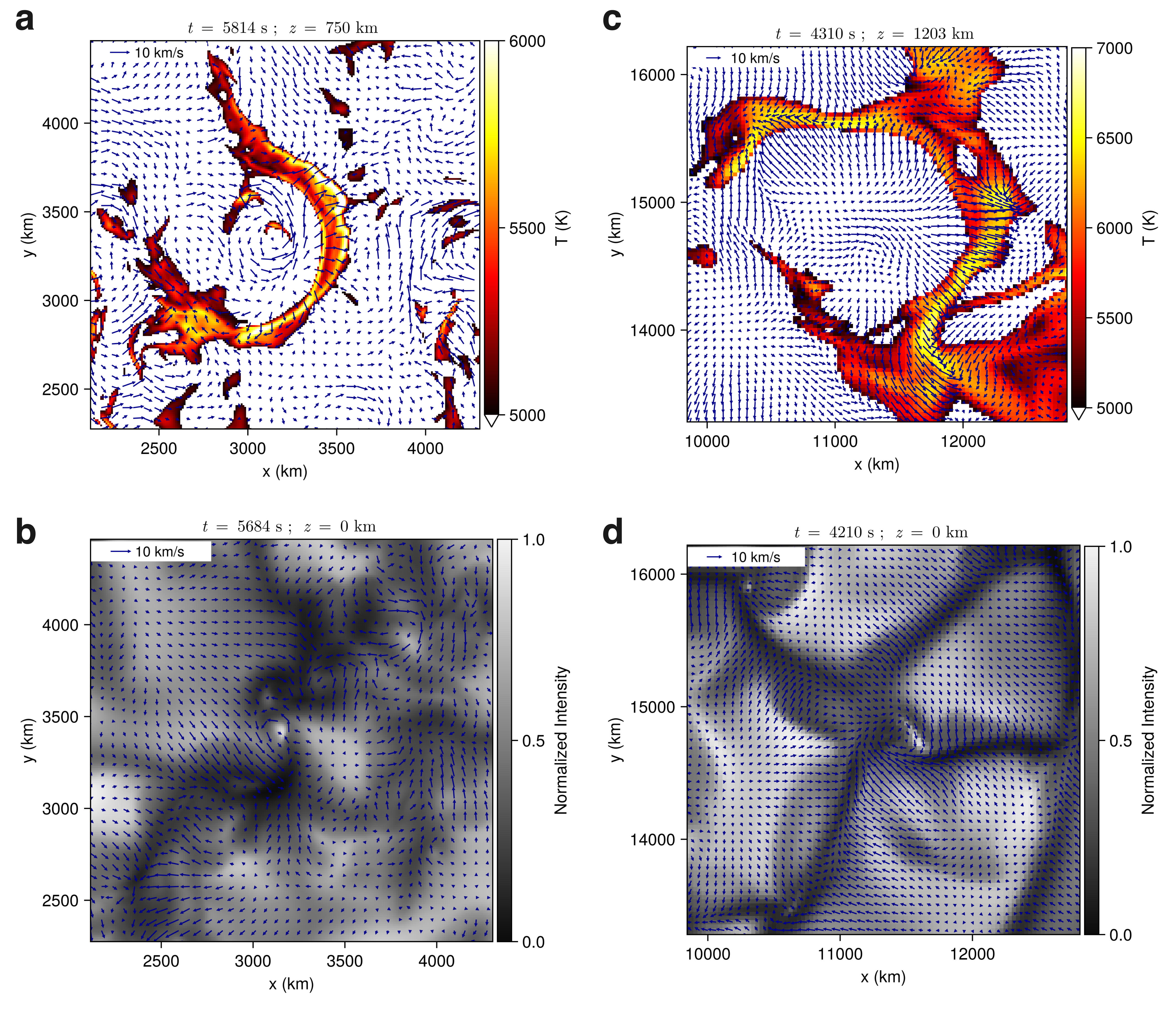}
	\caption{Horizontal velocities from the simulated atmosphere. (a) The temperature above $5000$~K is shown as a background image to the horizontal velocity depicted with blue arrows for event Nr.~S1 (CO5BOLD) at a height of $z = 750$~km at time 
$t = 5814$~s. (b) Continuum intensity at a wavelength of $450$~nm as a background to the horizontal velocity for event Nr.~S1 at a height of $z = 0$~km at time $t = 5684$~s. (c) The temperature above $5000$~K is shown as a background image to the horizontal velocity depicted with blue arrows for event Nr.~S2 (Bifrost) at a height of $z = 1200$~km at time 
$t = 4310$~s. (d) Continuum intensity at $450$~nm as a background to the horizontal velocity for event Nr.~S2 at a height of $z = 0$~km at time $t = 4210$~s.}
	\label{fig:Sim_phot_chrom_hvel}
\end{figure*}
\begin{figure*} [h!]
	\centering
	\includegraphics[width=\textwidth]{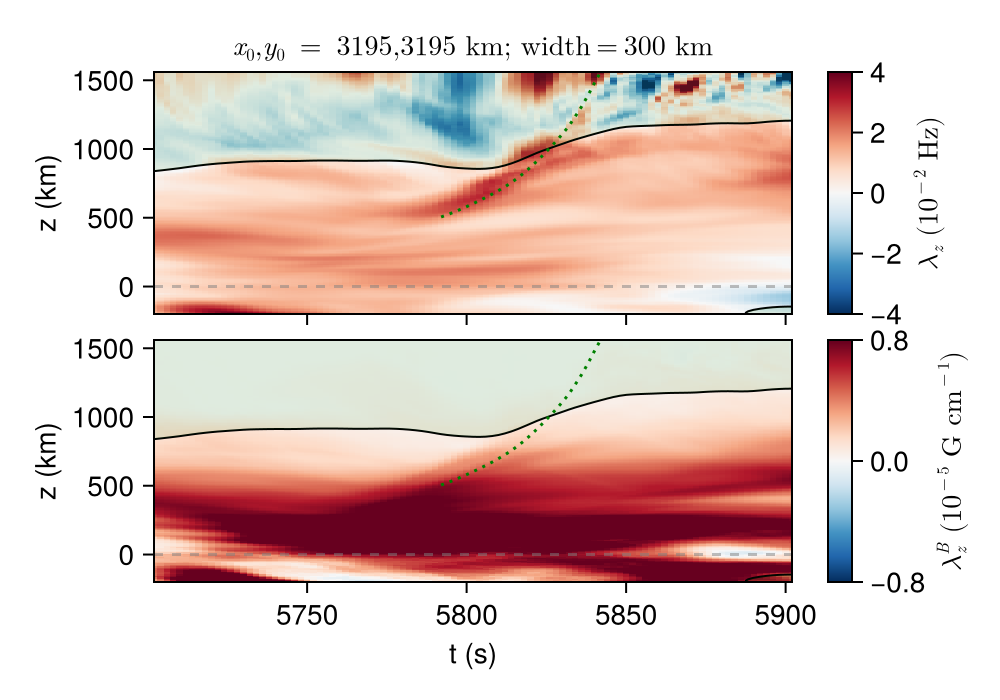}
	\caption{Time-distance diagrams of swirling strength and magnetic swirling strength. The top and bottom panels show the vertical component of the swirling vector $\lambda_{z}$, and the magnetic swirling vector $\lambda_{z}^{B}$, respectively as a function of height and time for simulation event Nr.~S1 shown in Figure~\ref{fig:Sim_ex1_overview_comp}. Swirling vector and magnetic swirling vector can be considered a generalisation of fluid vorticity and magnetic twist, respectively (see \cite{2021A&A...649A.121B}). The quantities are averaged over a horizontal area of $300 \times 300$~km$^2$ centerd on the event at coordinates $(x, y) = (3195, 3195)$~km. The dotted paths show for comparison the trajectory of a test particle moving with the Alfv{\'{e}}n speed plus vertical component of velocity. The speeds are averaged over the horizontal area. The light shaded part above the black curve shows the region where the average magnetic field is pointing upward; below it, it is pointing downwards. The ridges along the dotted curves have the same sign in $\lambda_{z}$ and $\lambda_{z}^{B}$ and move approximately with the Alfv{\'{e}}n speed, indicative of a torsional Alfv{\'{e}}n pulse that travels along the flux concentration with oppositely directed magnetic field. The ridges start to become visible in the top of the photosphere only where the magnetic field starts to become dynamically relevant. Also their appearance is sensitive on the precise location of the chosen line-of-sight and area due to the horizontal displacement of the vortex and magnetic field with time.}
	\label{fig:Sim_time_dist_swirl}
\end{figure*}

The simulation data show propagating swirls, which can be traced from the photosphere to the chromosphere, where they appear in close vicinity to the bright front. Figure~\ref{fig:Sim_phot_chrom_hvel} shows the horizontal component of the velocity field for simulation event Nr.~S1 in the left column of panels and for simulation event Nr.~S2 in the right column. For both events a swirl at the chromospheric level (top row) is seen to be in the location from where the bright front emanated. It is preceded by a swirl at the photospheric level (bottom row). The time delay between the photospheric swirl (shown in  Figure~\ref{fig:Sim_phot_chrom_hvel}(b) for the simulation event Nr.~S1) and the first appearance of the bright front in the synthetic \ion{Ca}{2}\,K maps at time $t = 5780$~s (visible from the animation linked to Figure~\ref{fig:Sim_ex1_tdev_vel}) is $96$~s for simulation case Nr.~S1 and $50$~s for simulation case Nr.~S2. For the case of simulation event Nr.~S1, Figure~\ref{fig:Sim_time_dist_swirl} shows the swirling strength and the magnetic swirling strength as a function of height
and time. Here, the swirling
strength and magnetic swirling strength are defined according to \citet{2021A&A...649A.121B}; they can be considered to be a measure of vortical motion and magnetic twist, respectively. From it, one can see the swirling motion propagating along the green dotted curve from the upper photosphere to the low chromosphere in phase with a propagating twist in the magnetic field, where the magnetic field points opposite to the propagation direction. This phase relation and polarity is indicative of a propagating torsional Alfv{\'e}n pulse in this region. More explanations are provided in the caption to Figure~\ref{fig:Sim_time_dist_swirl}. 

\section{Discussion} \label{sec:disc}

In this paper we report on the first DKIST VBI observation revealing chromospheric waves in the form of bright fronts in the spectral passband of \ion{Ca}{2}\,K that expose their source by manifesting a distinct arc shape. We detect at their apparent origin chromospheric bright grains and co-located G-band bright points, the typical signatures of photospheric magnetic flux concentrations. Furthermore, we find atmospheric vortices at the location of the G-band bright points prior to the launch of the chromospheric waves. Previous observational studies have inferred ubiquitous propagating Alfv{\'e}n waves or pulses from spectroscopic observations \citep{2009Sci...323.1582J, 2019NatCo..10.3504L}. However, these have not been reported to be connected to chromospheric traveling fronts as detected in our current study. Previously, a case of a radially traveling front has been reported much higher up in the transition region following a small-scale jet triggered within a magnetic element \citep{2011ApJ...743...14S}. However, this event was not associated with a vortex as in our study. 
\cite{2016A&A...586A..25P} found propagating arcs around a vortex region, which they said would become the subject of future work. No further details are given in the paper and it is not clear if this is the same phenomenon as studied in the present paper.
Also, previous analysis of simulation data have shown torsional Alfv{\'e}n pulses that propagate from the photosphere into the chromosphere very commonly to occur \citep{2021A&A...649A.121B,2024A&A...682A.181C} but again without the detection of a horizontally travelling bright front. 

The vortices detected in our study are capable of inducing torsional motions within the magnetic flux concentrations. Due to the magnetic flux concentrations exhibiting an inhomogeneous atmosphere, in particular torsional Alfv{\'e}n pulses 
can couple to other types of magnetohydrodynamic waves 
\citep{2021ApJ...911...39S,2007ApJ...671.1005B}. In particular, we find coupling to external, non-linear, predominantly acoustic waves in the ambient atmosphere that may steepen into shock fronts as the one seen in Figure~\ref{fig:Sim_ex1_overview_comp}(a). We note that the events bear a similarity to penumbral running waves identified as ``in effect low-$\beta$ slow-mode waves propagating along the magnetic field” \citep{2007ApJ...671.1005B}.  Although we do see a slow mode (field-aligned) component in the low-$\beta$ plasma region in simulation event S1 (above the black $\beta = 1$ contour in Figure~\ref{fig:Sim_time_distance_temp}), the shock front already appears in the high-$\beta$ region at about $5000$\,--\,$6000$ K (below the black $\beta = 1$ contour in Figure~\ref{fig:Sim_time_distance_temp}), hence stemming from a fast, predominantly acoustic wave.

Furthermore, while previous studies \citep{2009Sci...323.1582J,2019NatCo..10.3504L}
show that ubiquitous torsional Alfv{\'e}n pulses are present in the solar atmosphere, we now provide new evidence of vortical structures emitting magneto-sonic waves that shock and therefore dissipate in their vicinity. This is a major discovery of the present study, which is of general astrophysical interest since it points us to an unexpected route of dissipating torsional Alfv{\'e}n waves, which otherwise would hardly dissipate. It also may spur MHD laboratory experiments similar to those of \cite{PhysRevLett.127.275001} and is reminiscent of MHD technologies of hypersonic applications. 

\subsection{Energy estimates} \label{subsec:energy_est}
It also naturally raises the question of how energetic these events are. We estimate for simulation event Nr.~S1 that the shock event supplies an average power of $\sim2.46 \times 10^{15}$~W (see Appendix~\ref{append:energy_estimate} for further details). The required energy flux to sustain the lower chromosphere is estimated at $\sim2\times10^{15}$~W\,Mm$^{-2}$ (see in the book by \cite{2014masu.book.....P}, Table~1.1 on page 8). Assuming that a typical shock event provides the power found in the simulation and all of its energy is used for replenishing chromospheric radiative losses, $0.8$ per Mm$^2$ events would then be required to be constantly present to energize the lower chromosphere.
So far we found a much lower occurrence rate of $0.14$ events per Mm$^2$ per hour. Torsional Alfv{\'e}n pulses have been observed to occur with a rate of up to $0.08$ per Mm$^2$ at any time \citep{2022A&A...663A..94D} and simulations suggest a density of up to $\sim1$~Mm$^{-2}$ at any time \citep{2024A&A...682A.181C}, which indicates the potential of a much higher occurrence rate for swirl driven shock waves. However, taking the present results at face value, these events remain a contributor, not the main contributor to chromospheric heating.

\subsection{Contribution to the solar and stellar chromospheric basal flux} \label{subsec:basal}
In a broader context, more than $85$\% of the stars in our galaxy have surface hydrogen convective envelopes, like our own Sun, and exhibit solar-like activity and variability. There is increasing evidence that solar-like bright points \citep{2015A&A...581A..42B, 2018A&A...614A..78S} exist in these objects while magnetoacoustic waves are thought to provide a basal energy source for the heating of their chromospheres \citep{2011MNRAS.414..418P}. For providing the solar chromospheric H and K basal flux, an occurrence rate of $0.14$ per Mm$^2$ at any time would suffice (see Appendix~\ref{append:energy_estimate} for further details). This possibility is of particular interest for stellar physics with regard to the relative chromospheric H and K basal emission index as a function of colour index \citep{2018A&A...616A.108B, 2024A&A...689A.198C}.

\subsection{Future direction} \label{subsec:future}
 The observed events represent one of the rare observations of how the readily available kinetic energy present in the photosphere can be channelled into the outer regions of the solar atmosphere. Other diagnostics may reveal events similar to the ones presented in this paper but occurring in different layers of the solar atmosphere, possibly even in the corona. In consequence, they warrant further analysis such as a multiwavelength spectropolarimetric study to reveal their thermodynamic and magnetic structure.
Identifying the possible connectivity and modes of energy conversion throughout the solar atmosphere will lead to advances in deciphering the signal of unresolved stars. While we only catch glimpses of the processes and phenomena that arise from the surface of other stellar photospheres \citep{2011A&A...525A.140D, 2024MNRAS.527.7681L}, the DKIST observations reported here allow us to investigate the complex interplay between the plasma and the magnetic field at an unprecedented level of detail.

\begin{acknowledgments}
The research reported herein is based on data collected with the Daniel K. Inouye Solar Telescope (DKIST), a facility of the National Solar Observatory (NSO).  NSO is managed by the Association of Universities for Research in Astronomy, Inc., and it is funded by the National Science Foundation.  Any opinions, findings, and conclusions or recommendations expressed in this publication are those of the authors and do not necessarily reﬂect the views of the National Science Foundation or the Association of Universities for Research in Astronomy, Inc.  DKIST is located on land of spiritual and cultural significance to Native Hawaiian people.  The use of this important site to further scientific knowledge is done so with appreciation and respect.

The authors wish to acknowledge scientific discussions with the Waves in the Lower Solar Atmosphere (WaLSA; https://www.WaLSA.team) team, which has been supported by the Research Council of Norway (project no. 262622) and The Royal Society (award no. Hooke18b/SCTM). This research was also supported by the International Space Science Institute (ISSI) in Bern, through ISSI International Team project 502 (WaLSA: Waves in the Lower Solar Atmosphere at High Resolution). 

JRCC and OS acknowledge support by the Swiss National Science Foundation under grant IDs 200020\_182094 and 2000020\_15103 and the University of Zürich under grant UZH Candoc 2022 ID 7104.

OS acknowledges HPC resources by the Swiss National Supercomputing Center (CSCS) under project ID s560.

RE is grateful to STFC (UK, grant number ST/M000826/1) and acknowledges support from NKFIH Hungary (OTKA, grant No. K142987 and Excellence Grant, grant nr TKP2021-NKTA-64) and NSF China (PIFI, grant number No. 2024PVA0043). 

SJafarzadeh and DBJ wish to thank the UK Science and Technology Facilities Council (STFC) for the consolidated grants ST/T00021X/1 and ST/X000923/1.
DBJ also acknowledges support from the UK Space Agency for a National Space Technology Programme (NSTP) Technology for Space Science award (SSc 009), as well as The Leverhulme Trust for grant RPG-2019-371. 

MM acknowledges support from the Science and Technology Facilities Council (STFC) under grant No. ST/P000304/1, ST/T001437/1, ST/T00021X/1 and ST/X000923/1, as well as The Leverhulme Trust for grant RPG-2019-371.

Ivan Milić is gratefully acknowledged for his valuable assistance with the Lightweaver code.

\end{acknowledgments}


\clearpage
\appendix

\section{Differences between the two simulation data}
\label{append:diff_sim}
In-depth analysis of simulation and synthetic data has been primarily carried out on the CO5BOLD data. However, we also use a Bifrost simulation and synthetic data to consolidate the analysis. CO5BOLD data are used in the first place for its higher spatial resolution ($10$~km vs. $31.25$~km computational cell size) and lower diffusivities (see further below) compared to the available Bifrost data. The initial condition for the magnetic field of the two models is fundamentally different. In both cases, the initial condition for the magnetic field remains imprinted on the subsequent time evolution. CO5BOLD data started with a homogeneous vertical field with a strength of $50$~G (mimicking a predominantly unipolar magnetic network patch) while Bifrost data started with a unipolar vertical field of $5$~G, mimicking the mean signed field strength of a coronal hole, and was subsequently supplied with horizontal magnetic field with a strength of $200$~G that slowly entered across the bottom boundary in regions of upflows, leading to an averaged unsigned magnetic field strength in the photosphere of $40$~G. This situation results in a more turbulent, mixed polarity magnetic field at the solar surface than in the case of the CO5BOLD data. Details on the boundary conditions are given in \cite{2018Calvo_thesis} (Sect. 2 of the PhD Thesis) for the CO5BOLD data and in \cite{2016A&A...585A...4C} and under \url{https://sdc.uio.no/search/simulations} for the Bifrost data and in particular in \cite{2022A&A...665A.118F} for the Bifrost model ch024031\_by200bz005.

Regarding diffusivities, the two data sets differ as well. The CO5BOLD model features a shock capturing Riemann type numerical solver (HLL; \cite{1983Harten}) based on a reconstruction scheme \citep{2013MSAIS..24...26F} that keeps diffusivities locally minimal while preserving stability of the numerical solution. This renders, in particular, relatively sharp shock fronts. Bifrost, on the other hand, uses an explicit global and local-specific viscosity and resistivity, large enough to keep the solution stable in time. For the CO5BOLD data, we estimate an effective kinematic viscosity and plasma resistivity near the top of the photosphere of $\nu_{\rm eff} = 4.0\times10^{10}$~cm$^2$\,s$^{-1}$ and $\eta_{\rm eff} = 3.3 \times 10^{10}$~cm$^2$\,s$^{-1}$, respectively \citep{2022A&A...660A.115R}. For the Bifrost data, no such estimate is available but diffusivities are likely larger mainly, but not only, as a result of the lower spatial resolution. The Bifrost data includes a corona with Spitzer heat conductivity and a transition region to the chromosphere with optically thin radiative cooling (but H-ionization is treated in LTE). None of this physics is included in the CO5BOLD data, which reaches to a height of $1560$~km above optical depth $\tau_{\rm R} = 1$ only. Despite these significant differences in the production of the simulation data, we find in both the CO5BOLD and the Bifrost data events that much resemble the observed ones corroborating that these arise from a robust process not depending on the details of the initial data and boundary conditions.

\clearpage
\section{Additional observation examples}
\label{append:additional_examples}

The following figure shows sample images for all $10$ events detected in the observational data set.
\begin{figure*} [h!]
	\centering
	\includegraphics[width=\textwidth]{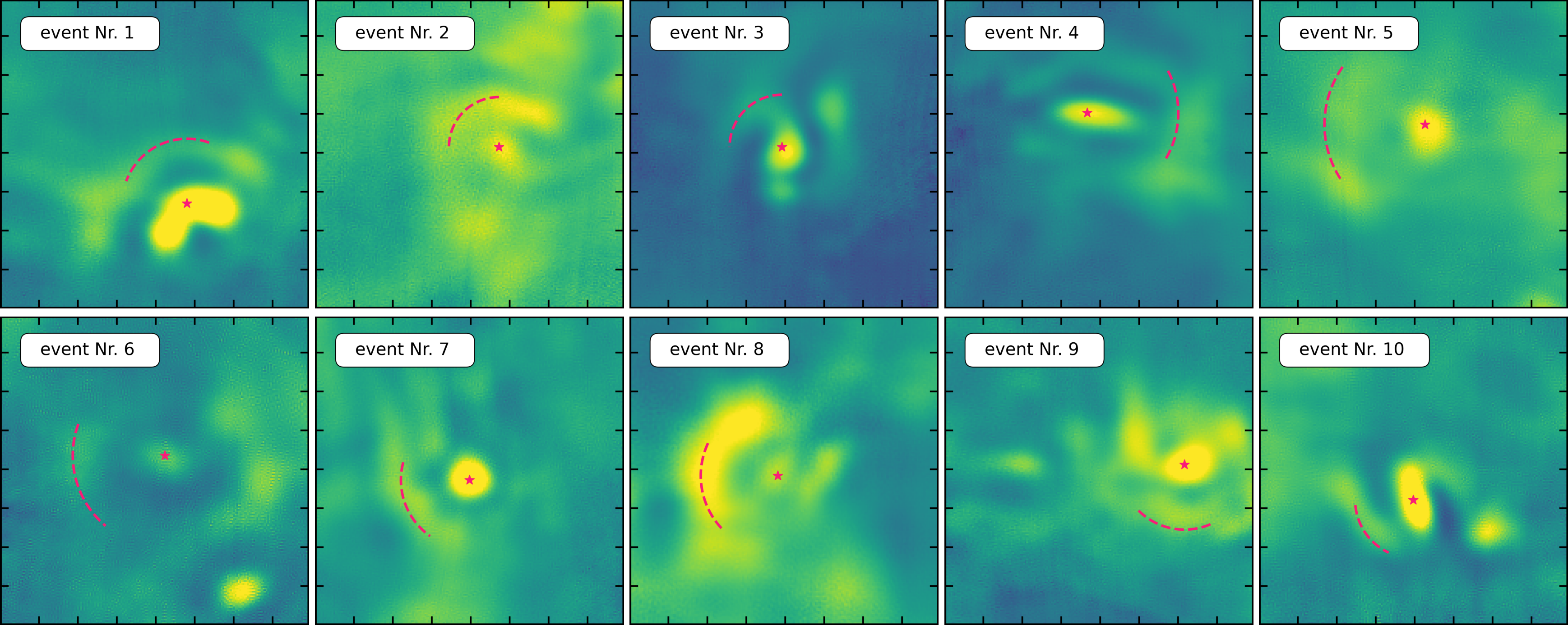}
	\caption{Collage of observed bright front events. The panels show time instances of the events seen in \ion{Ca}{2}\,K with tick marks set $200$~km apart. Part of the bright arc-shaped front is marked by a dashed pink arc. A pink star indicates the location of the \ion{Ca}{2}\,K bright grain and corresponding G-band bright points from where the bright front emanates. An animation showing the temporal sequence of all events is linked to the present Figure. The animation shows the continuum maps in the left panel and the \ion{Ca}{2}\,K maps in the right panel. In the time sequence of duration $1541$~s, each of the ten events is highlighted by a colored circle during its occurrence and numbered according to the event numbers given in this figure.}
	\label{fig:Obs_collage}
\end{figure*}

\clearpage
\section{Additional similar events in the simulation data}
\label{append:snthetic_imagery}

\begin{figure*} [h!]
	\centering
	\includegraphics[width=0.75\textwidth]{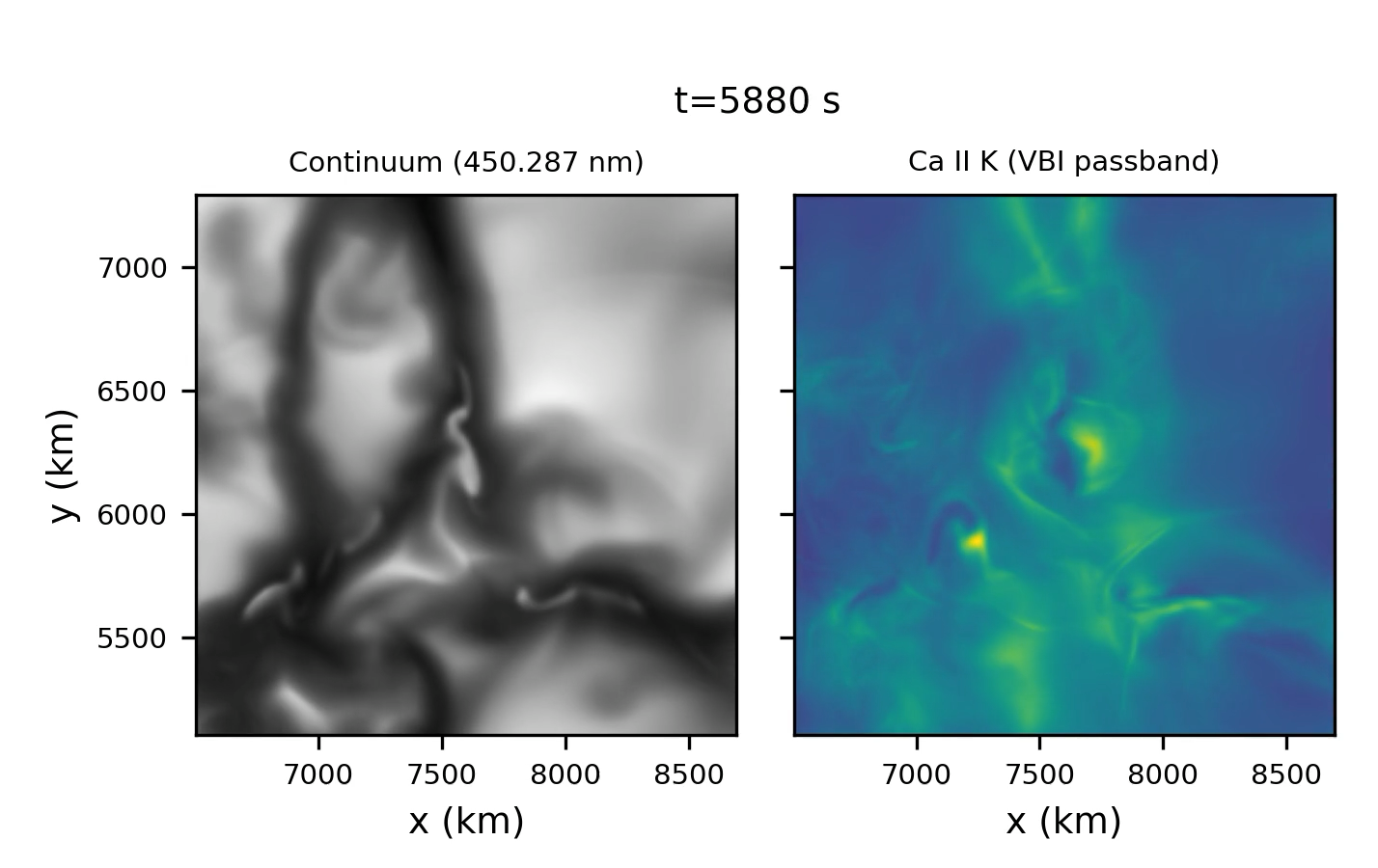}
	\caption{Synthesized maps from simulation event Nr.~S3: The animation shows in the right panel 
    the synthesized \ion{Ca}{2}\,K passband filter maps referring to the low chromosphere/upper photosphere and the left panel 
    maps of the continuum intensity at $450$~nm as a close proxy for the observational G-band images, both computed from the CO5BOLD numerical simulation. The time sequence of duration $102$~s shows the bright arc-shaped front marked in the \ion{Ca}{2}\,K map at times $t = 5896$\,s, $t = 5904$\,s, and $t = 5912$\,s by a dashed pink arc in the replay. The pink star indicates the location of the photospheric bright point during the time span of the event.}
	\label{fig:Sim_ex3_tdev}
\end{figure*}

We identify through visual inspection of a $632$~s long CO5BOLD high-cadence sequence sixteen events showing a scenario similar to the above-described cases Nrs.~S1 and S2, which all are associated with a swirl as well as with a photospheric magnetic flux concentration. We find, similar to the observations, apart from the basic common appearance, a variety in duration, shape, and life-time of the bright fronts. Figure~\ref{fig:Sim_ex3_tdev} shows a time instance and corresponding animation of simulation event Nr.~S3 from the CO5BOLD data. The animation linked to Figure~\ref{fig:Sim_ex3_tdev} depicts, as an example, an event with a wave front that does not exhibit a large apparent horizontal propagation speed. We found one more event with a shock-front topography very similar to simulation case Nr.~S1; in the remaining cases, the unambiguous tracking of individual
shock fronts was challenging due to the presence of multiple temperature peaks and wave activity from various other sources. The number of found events is, as in the observations, most likely a lower limit for these types of events due to the likelihood of events going unnoticed during visual inspection or forming in a different height range. The occurrence rate is by a factor of about seven larger in the simulations compared to the observations which are indeed susceptible to data quality and seeing limitations.

\clearpage
\section{Wave/shock identification and energy estimate}\label{append:energy_estimate}
For simulation event S1, the shock front is identified as a sharp temperature enhanced front in the computational domain (see Figure~\ref{fig:Sim_ex1_overview_comp} and Figure~\ref{fig:Sim_time_distance_temp}). We then construct a surface in the three-dimensional space based on this peak in temperature that identifies the shock front at different times. 

\begin{figure*} [h!]
	\centering
	\includegraphics[width=\textwidth]{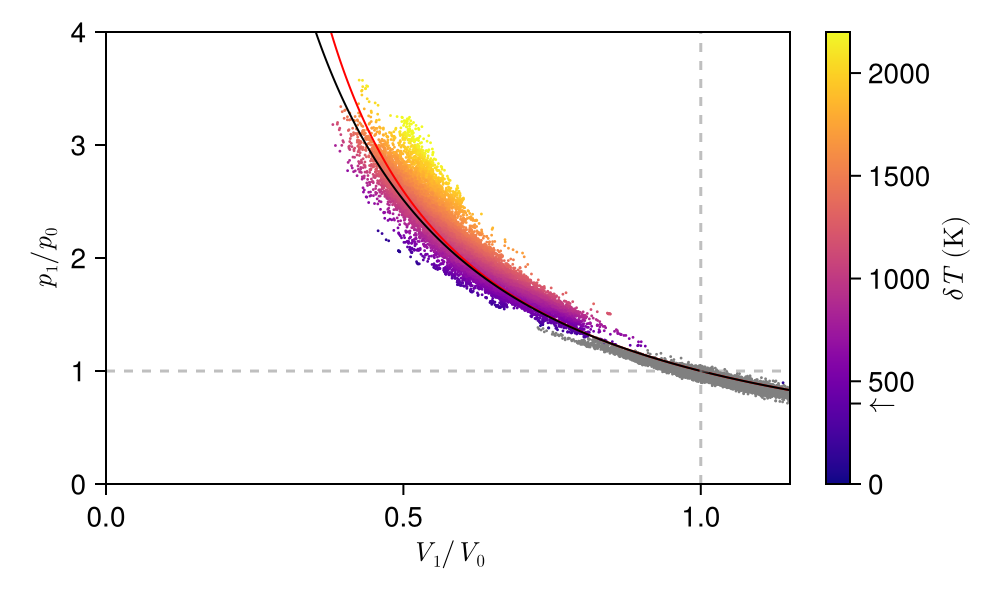}
	\caption{Pressure-Volume ($p$\,-$V$) diagram showing the shocked and the quiescent condition. The temperature difference between pre- and post-shock medium is displayed. The quiescent conditions (grey dots) follow the isentropic curve with minor volume compression and expansion typical of an acoustic disturbance. The temperature variations in the quiescent conditions are less than $392$~K (marked by an arrow in the colourbar). The shocked state (coloured dots) still essentially follows the isentropic curve but shows larger temperature changes with compression and pressure enhancements. Yet, plasma speeds relative to the shock front stay close to sonic, indicative of a weak shock. For comparison, the red and black curves show the Hugoniot and isentropic tracks, respectively, for a constant, mean $\gamma = 1.33$. }
	\label{fig:Sim_press_volume}
\end{figure*}
\begin{figure*} [h!]
	\centering
	\includegraphics[width=14cm]{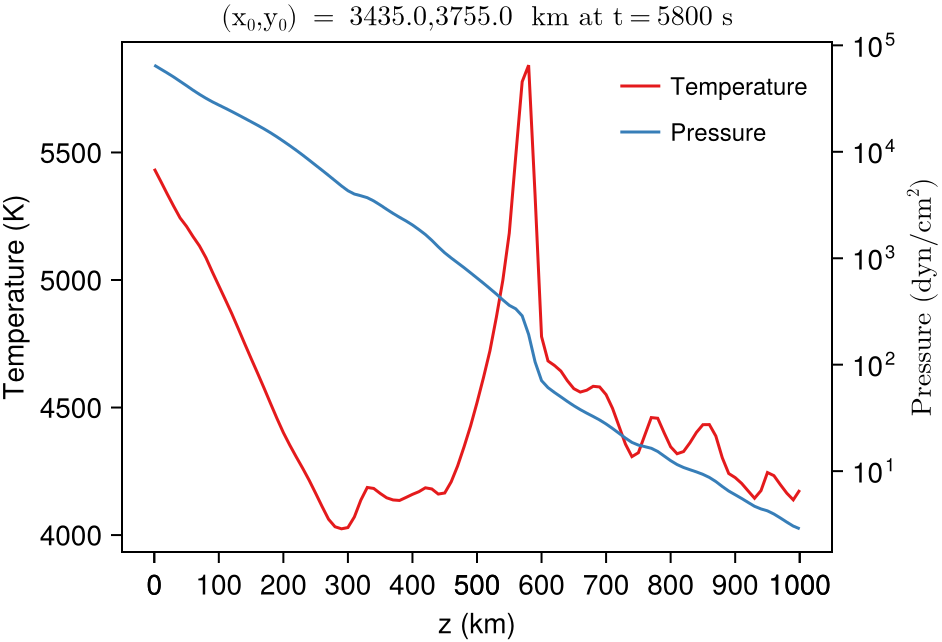}
	\caption{Temperature and gas pressure across the shock front.
Temperature (red) and gas pressure (blue) along the white line-of-sight of Figure~\ref{fig:Sim_ex1_overview_comp} at $(x, y) = (3435~\mathrm{km}, 3755~\mathrm{km})$ and time $t = 5800$~s.}
	\label{fig:Shockfront}
\end{figure*}

Regarding the wave properties, Figure~\ref{fig:Sim_press_volume} shows the pressure-volume ($p$\,-$V$) diagram for the shocked state and a quiescent condition for the event presented in Figure~\ref{fig:Sim_ex1_overview_comp}. For each snapshot ($t = 5812$~s to $5820$~s), we extract the thermodynamic variables pressure, density, and temperature on the shock surface ($p_1$, $\rho_{1}$, $T_1$) while the pre-shock values ($p_0$, $\rho_{0}$, $T_0$) are extracted 2 seconds prior to the arrival of the shock on the same surface. For characterization of the quiescent condition, we use the thermodynamic variables at 2 and 4 seconds before the arrival of the shock at the surface. In any case, the simulation data take a realistic equation-of-state with variable adiabatic index into account. The variable $V$ is the specific Volume $1/\rho$. The quiescent condition, marked as grey dots in Figure~\ref{fig:Sim_press_volume}, follow closely the black isentropic ($p\,V^{\gamma} =$ constant) curve with minor volume compression/expansion typical of acoustic disturbances and temperature fluctuations that do not exceed $392$~K. The shocked state (coloured dots) still essentially follows the isentropic curve but deviates from the quiescent state in that it shows temperature changes of up to $2429$~K with compression and a pressure enhancement of up to $3.5$ times the pre-shock value. Yet, the pre-shock speeds relative to the shock front stay close to sonic, indicative of a weak shock \citep{1966pswh.book.....Z}. The shock wave starts in a high $\beta$ environment as a predominantly acoustic shock at t $\approx 5790$~s (see Figure~\ref{fig:Sim_time_distance_temp}) but continues as a slow mode, still predominantly acoustic, in the time frame $5812$~s to $5820$~s. Therefore, we can compare it with results from pure hydrodynamics.  The spread in the shocked state is due to the spread in the adiabatic index $\gamma$ in the shocked media. In the case of a magnetic field-free, perfect gas with constant adiabatic index $\gamma$, the shocked state will lie on the Hugoniot curve given as 
%
\begin{equation}
  \quad \quad
  \frac{p_1}{p_0} = \left(\displaystyle\frac{\gamma+1}{\gamma-1} - \frac{V_1}{V_0}\right)\left/\left(\displaystyle\frac{\gamma+1}{\gamma-1} \frac{V_1}{V_0} - 1\right)\right. 
\end{equation}
from Eq. 1.75 of  \cite{1966pswh.book.....Z}. The shock Hugoniot (red curve) closely follows the isentropic curve (black) for pressure changes of up to a factor 4 and starts deviating from the isentropic case for strong shocks only, which we do not have in the present situation. The close agreement of the numerical results (colored and grey dots) with the isentropic and the Hugoniot curve of an ideal, non-magnetic fluid confirms the acoustic character of the wave and the difference in temperature and pressure fluctuations before and across the wave-front its character as a weak shock. For a representation of the shock structure, Figure~\ref{fig:Shockfront} shows the jump in temperature and gas pressure across the shock front along the white line-of-sight of Figure~\ref{fig:Sim_ex1_overview_comp} at $(x, y) = (3435~\mathrm{km}, 3755~\mathrm{km})$ and time $t = 5800$~s. It shows the typical saw-tooth shape and the signature of a radiative precursor and relaxation of a radiative shock in a stratified atmosphere as discussed in the book by \citet[][chapter~8.2]{1984frh..book.....M} or by \citet{1997ApJ...481..500C}.

The wave energy-flux is estimated on 
the constructed shock surface as $\mathbf{F} = \Delta p\,\mathbf{v}$, where $\Delta p$ is the pressure perturbation relative to the background undisturbed medium 2~s prior to the arrival of the shock, and $\mathbf{v}$ is the velocity on the shock surface. The energy flux $\mathbf{F}$ is integrated over the area of the shock surface to estimate the power of the event quoted in Sect.~\ref{subsec:energy_est}.

We estimate the contribution to the chromospheric emission index $R^{'}_{\rm HK} = \log(F_{\rm HK}/F_{\rm bol})$, where $F_{\rm HK}$ is the radiative flux in the cores of the \ion{Ca}{2}\,H and K spectral lines and $F_{\rm bol} = \sigma T_{\rm eff}^{4}$, the bolometric flux, where $\sigma = 5.7 \times 10^{-5}$~erg\,cm$^{-2}$\,s$^{-1}$~K$^{-4}$ is the Stefan–Boltzmann constant and $T_{\rm eff}$ is the effective temperature, here $5778$~K. For simulation event Nr.~S1, we estimate that the shock front supplies a mean energy flux of $F_{\rm shock} = 1.12 \pm 0.16 \times 10^7$~erg\,cm$^{-2}$\,s$^{-1}$ for about $8$~s over a mean shock front area of $A = 2.52 \pm 0.30 \times 10^{15}$~cm$^2$. Assuming that all this energy flux converts into H and K emission, and that the solar surface is at any time occupied with such shock surfaces with a filling factor $f$, we can compute $R^{' \,\rm shock}_{\rm HK}$ = $\log(f\,F_{\rm shock}/F_{\rm bol} ) = \log f - 3.748$. The chromospheric basal flux of the Sun is $R^{'}_{\rm HK} = -5.2$ so that a filling factor of $\log f = -5.2 + 3.75 = -1.45$, hence $f = 0.035$, would suffice to supply the chromospheric basal flux by these shock events alone. Using the mean area $A$ of shock event Nr.~S1, this translates to one such event per $A/f = 7.2$~Mm$^2$ or an occurrence rate of $0.14$ per Mm$^2$. Derivation of $F_{\rm HK}$ from the DKIST \ion{Ca}{2}\,K filtergrams alone is conceivable but presently not possible. This is due to calibration issues, whereby the dimensional fluxes in erg\,str$^{-1}$\,s$^{-1}$ are not available.


\bibliography{DKIST_CAIIK_shocks}{}
\bibliographystyle{aasjournalv7}

\end{document}